\documentclass[aps,prx,twocolumn,showpacs,superscriptaddress,floatfix,10pt]{revtex4-1}

\usepackage{amssymb}  
\usepackage{color,soul} 
\usepackage{graphicx}  
\usepackage[tight]{subfigure}  
\usepackage{mathrsfs}
\usepackage[pdftex,pdfusetitle,bookmarks=true,colorlinks=true,citecolor=blue,urlcolor=blue,linkcolor=magenta]{hyperref}
\usepackage{hypcap}
\usepackage{braket}
\usepackage{xfrac}
\usepackage{tikz}
\usepackage{url}
\usepackage{setspace}
\usepackage{enumerate}
\usepackage{bm}
\usepackage{epstopdf}
\usepackage{amsmath}
\usepackage{amsfonts}
\usepackage{mhchem} 
\usepackage[normalem]{ulem}
\usepackage{tikz-feynman}

\DeclareMathOperator{\sgn}{sgn}

\newcommand{\be}{\begin{equation}}
\newcommand{\ee}{\end{equation}}
\newcommand{\bea}{\begin{eqnarray}}
\newcommand{\eea}{\end{eqnarray}}

\begin{document}

\title{Majorana end states in an interacting quantum wire}
\author{Ren-Bo Wang}
\affiliation{Department of Physics and Astronomy, University of Utah, Salt Lake City, Utah 84112, USA}
\author{Akira Furusaki}
\affiliation{Condensed Matter Theory Laboratory, RIKEN, Wako, Saitama 351-0198, Japan}
\affiliation{RIKEN Center for Emergent Matter Science, Wako, Saitama 351-0198, Japan}
\author{Oleg A. Starykh}
\affiliation{Department of Physics and Astronomy, University of Utah, Salt Lake City, Utah 84112, USA}

\begin{abstract} 
\label{abstract}
We propose and investigate a simple one-dimensional model for a single-channel quantum wire hosting electrons that interact repulsively and are subject to a significant spin-orbit interaction. We show that an external Zeeman magnetic field, applied at the right angle to the Rashba spin-orbit axis, drives the wire into a correlated spin-density wave state with gapped spin and gapless charge excitations. 
By computing the ground-state degeneracies of the model with either (anti-)periodic or open boundary conditions, we conclude that the correlated spin-density state realizes a gapless symmetry-protected topological phase, as the ground state is unique in the ring geometry while it is two-fold degenerate in the wire with open boundaries. Microscopically the two-fold degeneracy is found to be protected by the conservation of the {\em magnetization parity}.  Open boundaries induce localized zero-energy (midgap) states which are described, at the special Luther-Emery point of the model, by Majorana fermions. We find that spin densities at the open ends of the wire exhibit unusual long-ranged correlations despite the fact that all correlations in the bulk of the wire decay in a power-law or exponential fashion. Our study exposes the crucial importance of the long-ranged string operator needed to implement the correct commutation relations between spin densities at different points in the wire.
Along the way we rederive the low-energy theory of Galilean-invariant electron systems in terms of current operators. 
\end{abstract}
\date{\today}

\maketitle

\section{Introduction}
\label{sec:intro}
The search for condensed matter realization of Majorana fermions has been at the center of intense theoretical and experimental efforts in the last decade. It is strongly motivated by the promise of topological quantum computing as well as by its fundamental importance to our current understanding of numerous topological phases of matter \cite{Kitaev2001,Alicea2012}. Topological superconducting wires represent one of the most promising platforms for realizing Majorana end states \cite{Sau2010,Alicea2010,Lutchyn2010,Oreg2010}. By now, several experimental groups have reported transport and STM tunneling data consistent with Majorana physics \cite{Mourik2012,Nadj-Perge2014,Jack2019,Frolov2019,Loss2019}  and many more studies are currently under way. 

A topological superconducting wire is obtained by bringing a semiconducting quantum wire with significant spin-orbit interaction into close contact with an $s$-wave superconductor and then applying an external (Zeeman) magnetic field in the direction orthogonal to the Rashba spin-orbit axis of the wire \cite{Lutchyn2010,Oreg2010,Sato2010}. Provided that the chemical potential lies within the gap induced by the Zeeman field, the wire effectively realizes a one-dimensional $p$-wave superconductor which features localized Majorana states at the open ends of the wire (more generally, at the boundaries between topological and trivial phases). A single-channel topological superconducting wire has been generalized to more 
complex/other geometries such as multi-channel wires \cite{potter2010,brouwer2012}, wires with periodic modulation of the spin-orbit potential \cite{malard2016}, and chains of magnetic adatoms on the surface of a superconductor \cite{bernevig2013,glazman2015}. Realistic modeling of semiconductor-superconductor heterostructures has been developed \cite{loss2013,winkler2019}.

Electron interactions are very important in zero- and one-dimensional systems \cite{giamarchi2003,gogolin2004} and their effect on the topological properties of the suggested quantum wire setup were investigated early on \cite{fidkowski2011a,stoudenmire2011,gangadharaiah2011}, and some exact/rigorous results were obtained \cite{katsura2015,lapa2020}. 
In parallel, a search for strongly interacting wires with {\em algebraic} superconducting correlations \cite{fidkowski2011} which would remove the need for close proximity of the wire to the macroscopic superconductor has begun. 
Kitaev's toy $p$-wave superconductor model is characterized by the two-fold degeneracy of the ground states with {\em different fermion parities}, i.e., between the ground states with an even and an odd number of fermions in the wire with open ends \cite{Kitaev2001,turner2011}. 
This degeneracy makes it clear that in a wire with a fixed total number of electrons the conservation of the {\em subband parity} acquires crucial importance. 
One-dimensional models with superconducting inter-band interactions conserving subband parity \cite{fidkowski2011,sau2011,Cheng2011,keselman2015,keselman2018,Pasnoori2020} are found to possess two-fold degeneracy in their ground state and thus represent one-dimensional topological states with exponentially localized Majorana modes at their open ends. 
Interestingly, their topological nature is preserved despite the presence of the critical center-of-mass fluctuations in the bulk.

Subband parity, in the form of a specific {\em magnetization parity}, plays the key role in our work as well. 
Unlike most of the previous studies, however, we present a physical realization of the parity-conserving system in a realistic quantum wire with purely {\em repulsive} electron-electron interaction. 
Our key finding is that localized Majorana end states can be realized in a simpler setting which does not require proximity to an $s$-wave superconductor. 
All that is needed is a single-channel quantum wire with significant spin-orbit and strong repulsive interactions between electrons. 
Applying an external magnetic field in the direction orthogonal to the spin-orbit axis of the wire drives its many-electron state into a correlated spin-density wave (SDW) phase with a finite spin gap in the bulk and non-trivial magnetic correlations \cite{giamarchi1988} and transport properties \cite{Sun2007,suhas2008}. 
No superconductivity or fine-tuning of the chemical potential is required. We show below that these Majorana zero-energy states live in the particle-hole sector of the many-body problem and can be thought of as spin density operators localized near the wire's ends. 

Our manuscript is rather technical and is based on the bosonization technique as developed in Refs.~\onlinecite{Eggert1992,Wong1994,fabrizio1995,delft1998} and designed to account for the periodic, anti-periodic and open boundary conditions. 
It is organized as follows. In Sec.~\ref{sec:H} we formulate the Hamiltonian of the problem and show, with the help of renormalization group (RG) arguments, that the wire flows to strong coupling describing an interesting correlated SDW state. 
Focusing first on the wire in the ring geometry, which depending on the parity of the magnetization corresponds to either periodic boundary condition (PBC) or anti-periodic boundary condition (anti-PBC), we show that its ground state is unique. 
In Sec.~\ref{sec:obc} we consider the wire with open boundary conditions (OBC) and derive its effective fermion Hamiltonian at the special Luther-Emery point. 
The effective Hamiltonian is solved in Sec.~\ref{sec:refermion}, where we find that the ground state is two-fold degenerate. 
The physical importance of the magnetization parity and the physical manifestations of the discovered Majorana modes are analyzed in Sec.~\ref{sec:maj}. Our findings and physical insights derived from them are summarized in Sec.~\ref{sec:disc}. Numerous technical details of our calculations are described in three extended Appendices.

\section{Hamiltonian}
\label{sec:H}

We consider a single-channel quantum wire, the Hamiltonian of which consists of three main contributions, 
${\cal H}_{\rm wire} = {\cal H}_{\rm e} + {\cal V}_{\rm so} + {\cal V}_{\rm z}$. Here ${\cal H}_{\rm e} = {\cal H}_{\rm e,0} + {\cal H}_{\rm int}$ describes an ideal quantum wire,
\begin{subequations}
\begin{eqnarray}
{\cal H}_{\rm e,0} &=& \sum_s \int\! dx\,
 \Psi^\dagger_s(x) \!\left(-\frac{\partial_x^2}{2m} - \mu\right) \!\Psi_s(x),
\label{eq:1} \\
{\cal H}_{\rm int} &=& \frac{1}{2} \sum_{s,s'}\! \int\! dx dx'\,
 U(x-x') \Psi^\dagger_s(x) \Psi^\dagger_{s'}(x') \Psi_{s'}(x') \Psi_s(x),
\nonumber\\&&
\end{eqnarray}
\end{subequations}
where ${\cal H}_{\rm e,0}$ is the kinetic energy with chemical potential $\mu$, $U(x)$ is the screened Coulomb electron-electron (e-e) interaction, and $s=\uparrow,\downarrow$ is the spin index. The electrons are perturbed by the spin-orbit interaction 
\begin{equation}
{\cal V}_{\rm so} = \sum_{s,s'} \int dx\, \Psi^\dagger_s(x) (-i \alpha_R \sigma^y_{ss'} \partial_x) \Psi_{s'}(x)
\label{eq:2}
\end{equation}
as well as Zeeman magnetic field which we take to be directed along
the $\hat{z}$ axis, $\vec{B} = B \hat{z}$,
\begin{equation}
{\cal V}_{\rm z} = \sum_{s,s'} \int dx\, \Psi^\dagger_s(x) \left(-\frac{g \mu_B^{}}{2} \sigma^z_{ss'} B\right) \Psi_{s'}(x).
\label{eq:3}
\end{equation}
The spin-orbit interaction \eqref{eq:2} is obtained from the standard Rashba interaction, $\alpha_R \hat{z}\cdot\vec{p}\times\vec{\sigma}$, by replacing the transverse component of the electron momentum $\vec{p}$ by its zero expectation value, $p_y \to \langle p_y \rangle =0$. Corrections to this approximations are known to be very small \cite{Sun2007,suhas2008}. 

\begin{figure}
       \includegraphics[width=0.85\columnwidth]{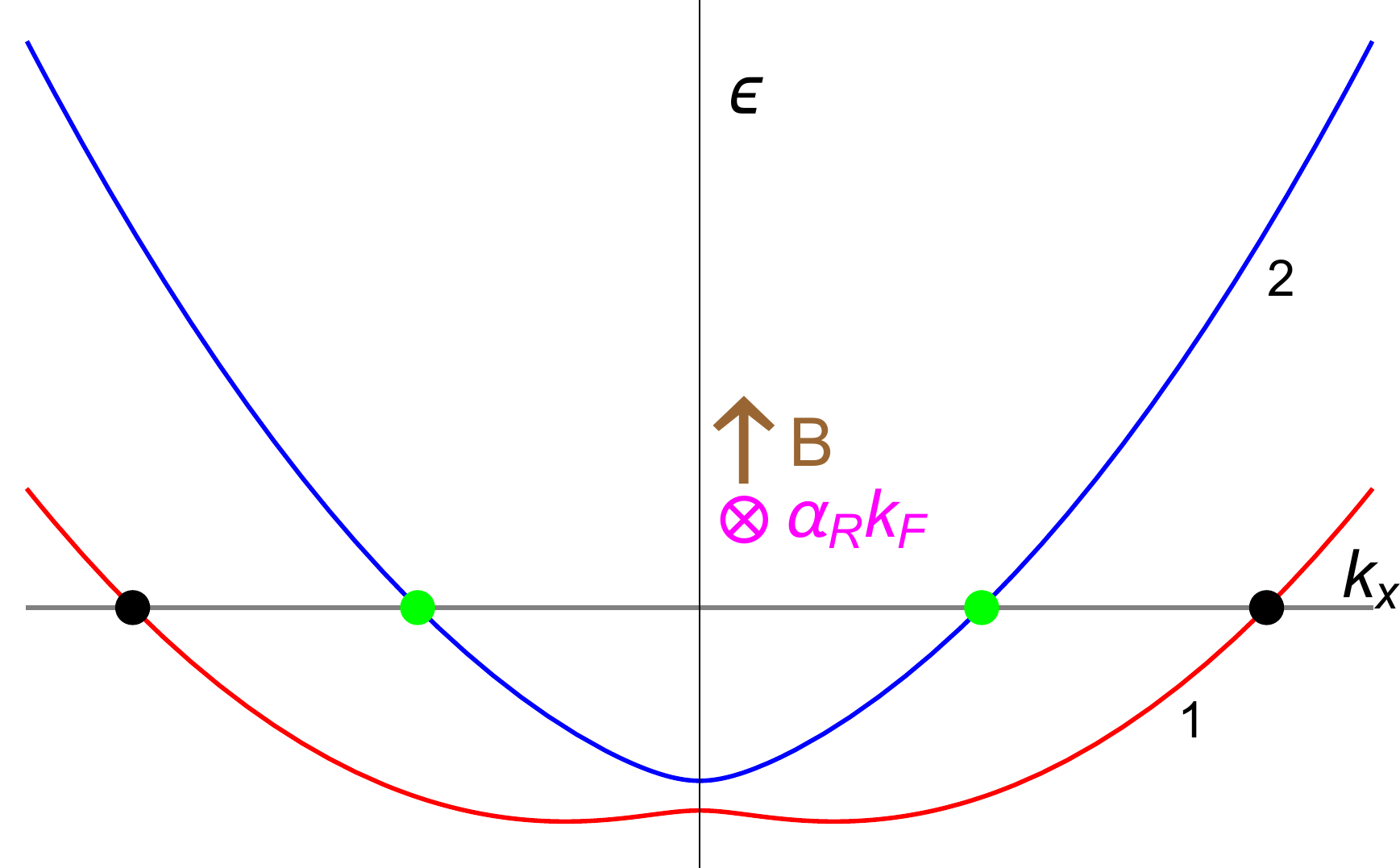}
         \caption{Schematics of the Cooper scattering process. Spin-orbit interaction is directed along the $\hat{y}$-axis, Zeeman magnetic field is applied along the $\hat{z}$-axis. 
         Two electrons in, say, the subband $1$ (represented by black points) are scattered into the opposite Fermi points in the subband $2$ (shown by green points). 
         In the the conjugate process two electrons initially in the band $2$ are scattered into the band $1$.}
     \label{fig:cooper}
\end{figure}

Obviously ${\cal H}_{\rm wire}$ does not conserve spin -- this fact is of crucial importance for our investigation. The key consequence of this can be understood by considering a limit of strong magnetic field $\mu \gg b \equiv g \mu_B^{} B \gg 2\alpha_R k_F$, where $k_F$ denotes the Fermi-momentum of the unperturbed Hamiltonian ${\cal H}_{\rm e,0}$, Eq.~\eqref{eq:1}. In this limit ${\cal H}_{\rm wire}$ describes the standard problem of a partially magnetized quantum wire with two Zeeman-split subbands. In the absence of the spin-orbit interaction no scattering  processes between these subbands are possible, simply because their spin wave functions are described by the orthogonal spinors, spin up ($\uparrow$) and spin down ($\downarrow$) states. This is just the consequence of the spin conservation. However, any finite spin-orbit interaction $\alpha_R \neq 0$ breaks spin conservation and immediately allows for a new scattering process: the Cooper scattering \cite{starykh2000}. This momentum- and energy-conserving process describes scattering of the pair of electrons at $\pm k_F$ Fermi points of, say, majority subband ($\uparrow$) into a similar pair of electrons in the minority ($\downarrow$) subband, and vice versa, see Figure~\ref{fig:cooper}. That is, a pair of electrons with spin $S^z = +1$ is converted into that with spin $S^z = -1$ and vice versa. This superconducting, or Josephson-like scattering (hence the name Cooper) conserves fermion parity of each of the subbands and plays a crucial role in the following discussion. 

Such a two-subband description can be straightforwardly extended to the physically most important regime of comparable Zeeman and spin-orbit energies, $b \approx 2\alpha_R k_F$, see for example Ref.~\cite{suhas2008}. We, however, will follow a less cumbersome approach, based on the chiral rotation trick, as detailed below. The end result of these complimentary calculations is the same \cite{suhas2008}.

\subsection{Spin current formulation of the quantum wire with periodic boundary conditions}
\label{sec:currents}

\subsubsection{Chiral fermions}
\label{sec:chiral fermions}

Our approach to the problem consists in treating perturbations ${\cal V}_{\rm so}$ and ${\cal V}_{\rm z}$ on equal footing.
Initially, we turn off the perturbations ${\cal V}_{\rm so}$ and ${\cal V}_{\rm z}$.
We express fermion fields $\Psi_s(x)$
in terms of low-energy modes $\Psi_{R s}$ and $\Psi_{L s}$ that live near $+k_F$ and $-k_F$ Fermi points, correspondingly,
\be
\Psi_s(x) = \Psi_{R s}(x) e^{i k_F x} + \Psi_{L s}(x) e^{-i k_F x}.
\label{eq:4}
\ee
The Fermi-momentum $k_F=\sqrt{2m\mu}$ is determined by
the electron density in the usual way, $k_F = \pi N^0_s/L$,
where $L$ is the length of the wire and $N^0_s = N^0_{R,s}+N^0_{L,s} = 2N^0_{R,s}$ is the total number of fermions of spin projection $s$.
It is written in terms of the numbers $N^0_{R/L,s}$ of chiral fermions in the wire.
The choice of $N^0_{R,s} = N^0_{L,s}$
made here corresponds to considering the state with \textit{no}
charge current in the ground state,
$j^0_{\rho}=\sum_s(N^0_{R,s} - N^0_{L,s}) = 0$.
In the absence of the external magnetic field the ground state
magnetization is zero as well, $M^0 = (N^0_\uparrow - N^0_\downarrow)/2 = 0$.

Consider the wire in the closed loop geometry, with the chiral
fermions \eqref{eq:4} obeying the periodic
boundary conditions (PBC) such that $\Psi_{R/L,s}(0) = \Psi_{R/L,s}(L)$.
Note that $e^{ik_FL} = 1$ due to our choice $k_F = 2\pi N^0_{R,s}/L$ explained below \eqref{eq:4}.

In terms of the chiral fermion fields $\Psi_{Rs}$ and $\Psi_{Ls}$, the kinetic energy is simply
\be
{\cal H}_{\rm e,0}  = \sum_s
\int dx \left( - i v_F \Psi_{R s}^\dagger \partial_x \Psi_{R s}
 + i v_F \Psi_{L s}^\dagger \partial_x \Psi_{L s}\right),
\label{eq:5}
\ee
where $v_F = k_F/m$ is the Fermi-velocity. 
It is useful at this stage to write the kinetic energy as a sum
of commuting {\sl charge} and {\sl spin} parts (Sugawara construction),
${\cal H}_{\rm e,0} = {\cal H}^0_{\rm \rho} + {\cal H}^0_{\rm \sigma}$, where
\bea
\label{H0-charge}
&&{\cal H}^0_{\rm \rho} = \frac{\pi v_F}{2} \int dx\, ( J_R^2 + J_L^2) ,\\
&&{\cal H}^0_{\rm \sigma} = \frac{2\pi v_F}{3} \sum_{a=x,y,z} \int dx\, ( J^a_R J^a_R + J^a_L J^a_L) .
\label{eq:6}
\eea
Here we introduced normal-ordered charge currents
\be
J_R = \sum_s :\!\Psi_{R s}^\dagger \Psi_{R s}\!: \, ,
\quad
J_L = \sum_s :\!\Psi_{L s}^\dagger \Psi_{L s}\!: \, ,
\label{eq:charge-current1}
\ee
and spin currents ($a=x,y,z$)
\be
J^a_R = \sum_{s,s'} :\! \Psi_{R s}^\dagger \frac{\sigma^a_{ss'}}{2}  \Psi_{R s'} \!: \, ,
\quad
J^a_L =  \sum_{s,s'} :\! \Psi_{L s}^\dagger \frac{\sigma^a_{ss'}}{2}  \Psi_{L s'} \!: \, .
\label{eq:spin-current1}
\ee

As described in Appendix \ref{sec: interaction and spin-charge separation}, the interaction part of the Hamiltonian \eqref{eq:1} separates into charge and spin parts as well, ${\cal H}_{\rm int} = {\cal H}_{\rm int, \rho} + {\cal H}_{\rm int, \sigma}$, where
\begin{subequations}
\bea
{\cal H}_{\rm int, \rho} &=&
\frac{1}{4}(2U_0-U_{2k_F})\int dx\,
(J_R + J_L)^2 ,
\label{eq:int-charge}
\\
{\cal H}_{\rm int, \sigma} &=& -g \int dx~  \vec{J}_R \cdot \vec{J}_L .
\label{eq:int-spin}
\eea
\end{subequations}
Here $U_q$ denotes $q$-th component of the Fourier transform of the e-e interaction $U(x)$, and $g = 2 U_{2k_F}$ denotes the magnitude of the spin backscattering interaction.

We now turn on the perturbations ${\cal V}_{\rm so}$ and ${\cal V}_{\rm z}$.
The Zeeman magnetic field $b = g \mu_B^{} B$ couples to the sum of spin currents (magnetization)
\be
{\cal V}_{\rm z}  = - b \int dx\, (J^z_R + J^z_L),
\label{eq:7}
\ee
while the spin-orbit interaction couples to their {\em difference}, since the Rashba term \eqref{eq:2} is odd under spatial inversion ($x \to -x$) which interchanges
right- and left-moving excitations \cite{Sun2007},
\be
{\cal V}_{\rm so} = 2\alpha_R k_F \int dx~ (J^y_R - J^y_L).
\label{eq:8}
\ee

\subsubsection{Chiral Rotations}
\label{sec:rotations}

It is crucial to notice now that ${\cal H}^0_{\rm \sigma}$ possesses an extended $SU(2) \times SU(2)$ symmetry of independent rotations of the right- and left-moving currents.
Our solution of the problem \cite{schnyder2008,suhas2008} exploits this extended symmetry. Namely, we next rotate $\vec{J}_R$ and $\vec{J}_L$ about the $\hat{x}$-axis in opposite directions so as to bring
``vectorial" perturbation  ${\cal V} = {\cal V}_{\rm so} + {\cal V}_{\rm z}$ into the standard Zeeman form, with {\em total} field $h= \sqrt{b^2 + (2\alpha_R k_F)^2}$
along the $\hat{z}$-axis
\be
{\cal V} = - h \int dx~ (M^z_R +  M^z_L).
\label{eq:9}
\ee
The required chiral rotation is given by 
\be
\vec{J}_R = {\cal R}_x(\beta_R) \vec{M}_R,
\quad
\vec{J}_L = {\cal R}_x(\beta_L) \vec{M}_L,
\label{Chiral Rotation}
\ee
where the rotation matrix ${\cal R}_x$ is
\bea 
{\cal R}_x(\beta) =
\left(
\begin{array}{ccc}
1 & 0 & 0 \\
0 & \cos\beta & -\sin\beta \\
0 & \sin\beta & \cos\beta \\
\end{array}
\right).
\label{eq:rot x}
\eea
The rotation angles are given by 
\be
\beta_R = -\beta_L = \beta= \arctan(2\alpha_R k_F/b).
\label{eq:a10 in Sec. II}
\ee
These rotations do not affect ${\cal H}_\sigma^0$ \eqref{eq:6}, which retains its form in the rotated $M$-basis
\be
{\cal H}^0_{\rm \sigma}
 = \frac{2\pi v_F}{3} \sum_{a=x,y,z} \int dx~  (M^a_R M^a_R + M^a_L M^a_L),
\label{H_sigma^0 M}
\ee
where $v_F$ here is understood as the one including the shift $-U_{2k_F}/3$ found in \eqref{H_s}.

In terms of the right- and left-moving fermions, the rotation \eqref{Chiral Rotation} corresponds to the rotation of spinors 
$\Psi_{R/L} = (\Psi_{R/L \uparrow}, \Psi_{R/L \downarrow})^T$
and $\psi_{R/L} = (\psi_{R/L \uparrow}, \psi_{R/L \downarrow})^T$,
\be
\Psi_{R} = e^{-i \beta \sigma^x/2} \psi_{R} ,
\quad
\Psi_{L} = e^{i \beta \sigma^x/2} \psi_{L}.
\label{rot-fermions}
\ee
This observation makes clear that the charge currents \eqref{eq:charge-current1} do not transform under the rotations \eqref{Chiral Rotation} and \eqref{rot-fermions} -- the Hamiltonian of the charge sector 
${\cal H}^0_\rho + {\cal H}_{\rm int, \rho}$ is not affected.
The new fermions parameterize the rotated currents $\vec{M}_{R/L}$
in the same way as the old ones parameterize the currents $\vec{J}_{R/L}$.
For example, under the right rotation ${\cal R}_x(\beta)$
\be
\vec{J}_R =
\, :\!\Psi^\dagger_{R} \frac{\vec{\sigma}}{2} \Psi_{R}\!:\, \to
\vec{M}_R =
\, :\!\psi^\dagger_{R} \frac{\vec{\sigma}}{2} \psi_{R}\!: .
\label{eq:11}
\ee

The interaction in the spin sector ${\cal H}_{\rm int, \sigma}$ \eqref{eq:int-spin}, is strongly modified by the rotation and changes to
\bea
{\cal H}_{\rm int, \sigma} &=&
 - g \int dx~ \vec{M}_R {\cal R}_x^T(\beta_R) {\cal R}_x^{}(\beta_L) \vec{M}_L \nonumber\\
&=& - g\int dx \left[
M^x_R M^x_L+
 \cos\chi (M^y_R M^y_L + M^z_R M^z_L)
\right.
\nonumber\\
&&\left.\qquad\qquad
+ \sin\chi (M^y_R M^z_L - M^z_R M^y_L) \right],
\label{Hint,sigma M}
\eea
where $\chi=\beta_R - \beta_L = 2\beta$ is the {\em relative} rotation angle.

Observe that the net field $h$, \eqref{eq:9}, points along the $\hat{z}$-axis.
The magnetic field induces incommensurate fluctuations in the system which make some of the terms in \eqref{Hint,sigma M} to oscillate fast with the coordinate. 
It is easy to see that $h$ can be absorbed into the kinetic energy of
fermions $\psi_{R/L}$ by a simple $x$-dependent transformation
\be
\psi_{R} \to e^{i t_\varphi x \sigma^z/2} \psi_{R},
\quad
\psi_{L} \to e^{-i t_\varphi x \sigma^z/2} \psi_{L},
\quad
t_\varphi = h/v_F.
\label{shifting fermi points}
\ee
As a result of this shift the transverse components
$M^x_{R/L} \pm i M^y_{R/L} = M^\pm_{R/L}$
of the rotated spin current acquire oscillating position-dependent factors,
$M^+_R \to M^+_R e^{-i t_\varphi x}$ and
$M^+_L \to M^+_L e^{i t_\varphi x}$. 
The immediate consequence of this is that many terms in ${\cal H}_{\rm int, \sigma}$ \eqref{Hint,sigma M} acquire $x$-dependent oscillations,
\bea
{\cal H}_{\rm int, \sigma} \!&=&\!
 - g \!\int\! dx \!\left\{
   \cos\chi M^z_R M^z_L 
 + \frac{\sin^2\frac{\chi}{2}}{2}
(M^+_R M^+_L + {\text{h.c.}}) \right.\nonumber\\
&&\qquad\quad
 + \frac{\cos^2\frac{\chi}{2}}{2}
(M^+_R M^-_L e^{- i 2t_\varphi x} + {\text{h.c.}}) \nonumber\\
&&\left.\qquad
 + i \frac{\sin\chi}{2}\!\left[
       (M^z_L M^-_R + M^z_R M^+_L) e^{i t_\varphi x} - {\text{h.c.}}\right]\! \right\}\! .
\nonumber\\
\label{Hint,sigma with oscillations}
\eea
Provided that the running backscattering coupling constant $g/v_F$ is small, all
oscillating terms, which represent momentum-nonconserving two-particle
scattering processes, average out to zero. Assuming this, we are allowed to drop all oscillating terms in \eqref{Hint,sigma with oscillations} and obtain the non-oscillating 
part of the spin-interaction Hamiltonian as 
\bea
{\cal H}_{\rm int, \sigma} 
&=& - \int dx \left[
g_c (M^x_R M^x_L - M^y_R M^y_L) + g_z M^z_R M^z_L
\right] \nonumber\\
&=& - \int dx \sum_{a=x,y,z} g_a M^a_R M^a_L ,
\label{eq:a3}
\eea
where
\bea
g_x &=& - g_y = g_c = g \frac{1-\cos\chi}{2}
=\frac{g(2\alpha_R k_F)^2}{b^2+(2\alpha_Rk_F)^2}, ~
\label{g_c(0)}
\\
g_z &=& g \cos\chi=g\frac{b^2-(2\alpha_R k_F)^2}{b^2+(2\alpha_R k_F)^2}.
\label{g_z(0)}
\eea

Note that at this point the complete Hamiltonian of the spin sector is given by the sum of equations \eqref{H_sigma^0 M} and \eqref{eq:a3}.
Importantly, the magnetic field is absent from it, it is absorbed into renormalization of the Fermi momenta $k_F \to k_{F s}$.
In fact, the coupling constants $g_a$ in \eqref{eq:a3} have implicit dependence on $h$ acquired through renormalization-group transformation from the original energy scale (of the order of the band width) to the effective magnetic field $h$.

The meaning of \eqref{shifting fermi points} is simple. 
It represents splitting of the Fermi-momentum $k_F$ into the spin-dependent ones $k_{F s} = k_F + s t_\varphi/2$. 
Given that $k_F$ is determined by the particle density, $k_F = \pi N_0/L$, the development of the spin-dependent Fermi momenta $k_{F s} = \pi N_s/L$
describes the appearance of the finite magnetization with $N_\uparrow > N_\downarrow$. 
Therefore, $\Delta k_F = t_\varphi/2 = \pi (N_\uparrow - N^0_\uparrow)/L$, so that $t_\varphi L = 2 \pi (N_\uparrow - N^0_\uparrow)=2\pi M$ is an integer multiple of $2\pi$,
since $N_\uparrow$ and $N^0_\uparrow$ are integers describing the number of spin-$\uparrow$ electrons in the system with finite $h$ and zero $h$, respectively.
The magnetization $M=(N_\uparrow-N_\downarrow)/2$ is also an integer.

After making the transformations, the fermions $\psi_{Rs}$ and $\psi_{Ls}$ obey the boundary conditions
\begin{subequations}
\label{BCs}
\bea
\psi_{R}(0)&=&e^{i\sigma_z t_\varphi L/2}\psi_{R}(L)
=(-1)^M\psi_{R}(L),
\\
\psi_{L}(0)&=&e^{-i\sigma_z t_\varphi L/2}\psi_{L}(L)
=(-1)^M\psi_{L}(L).
\eea
\end{subequations}
The boundary conditions depend on the parity of the magnetization $M$:
periodic for even $M$ and anti-periodic for odd $M$.
It is appropriate to note here that even though our subsequent
analysis will show that the magnetization $M$ is
not a conserved quantity in the ground state of the interacting
wire, the magnetization parity $(-1)^M$ is
conserved in the ground state.
Therefore the boundary condition \eqref{BCs} is well defined.

The anti-periodic boundary condition for odd $M$ can be implemented by introducing a magnetic flux threading the ring under the periodic boundary condition.
Thus we replace $\partial_x$ with $\partial_x-i\pi \lambda/L$ in the kinetic energy $\mathcal{H}_\mathrm{e,0}$ in \eqref{eq:5}, or equivalently we add
\be
\mathcal{H}_\mathrm{flux}=\frac{\pi \lambda v_F}{L}\int\!dx\, (J_L-J_R)
\ee
to the charge part of the kinetic energy $\mathcal{H}_\rho^0$.
Here we demand the integer parameter $\lambda$ to be
\be
\lambda=\left\{\begin{array}{ll}
0 & \mathrm{for}~(-1)^M=1,\\
1 & \mathrm{for}~(-1)^M=-1.
\end{array}\right.
\label{flux vs M}
\ee

\subsubsection{RG analysis}
\label{sec:rg}

Equations \eqref{H_sigma^0 M} and \eqref{eq:a3}
represent a non-trivial interacting problem, analysis of which requires renormalization group (RG) treatment. 
Let us assume for the moment that the wire length $L$ is large so that finite-size effects are negligible.
The couplings $g_a$ obey the famous Berezinskii-Kosterlitz-Thouless (BKT) RG flow \cite{giamarchi2003,gogolin2004},
\be
\frac{d g_x}{d \ell} = -\frac{g_y g_z}{2\pi v_F} , \quad
\frac{d g_y}{d \ell} = -\frac{g_x g_z}{2\pi v_F} , \quad
\frac{d g_z}{d \ell} = -\frac{g_x g_y}{2\pi v_F} ,
\label{RG equations}
\ee
where $\ell=\log(\alpha'/\alpha)$ describes increase of the short-distance cutoff from $\alpha$ to $\alpha'$.
As discussed in detail in \cite{chan2017}, the solution to the RG equations
\eqref{RG equations}
depends on the initial values [\eqref{g_c(0)} and \eqref{g_z(0)}] of the couplings involved.
Noting that $d(g_x^2 - g_y^2)/d\ell =0$ and the fact that for $\ell=0$ $g_x + g_y = 0$, we conclude that $g_x(\ell) = - g_y(\ell) = g_c(\ell)$ for all $\ell$.
Equations \eqref{RG equations} then reduce to the two coupled equations
\be
\frac{d g_z}{d \ell} = \frac{g_c^2}{2\pi v_F} ,
\qquad
\frac{d g_c}{d \ell} = \frac{g_c g_z}{2\pi v_F} ,
\label{two RG equations}
\ee
which too are characterized by the integral of motion $Y = g_z^2(\ell) - g_c^2(\ell)$.

It turns out that the solution is towards {\sl strong coupling}, meaning that
$g_{z,c}(\ell) \to +\infty$ for sufficiently large $\ell$, for all possible angles $\chi \in (0,\pi)$ \cite{chan2017}. 
This diverging solution implies an instability towards a correlated spin state with a nonvanishing excitation gap in the spin sector.
The spin gap can be estimated as $\Delta_\sigma \sim (v_F/\alpha) e^{-\ell_0}$ where $\ell_0$ is the RG scale
at which the dimensionless coupling constants diverge. The {\em minimal} value of $\ell_0$, corresponding to the strongest instability, occurs for $Y=0$. 
This corresponds to $\cos\chi = 1/3$ and implies $b = 2\sqrt{2} \alpha_R k_F$.
Therefore, the correlated spin state is strongest when Zeeman energy is comparable to the spin-orbit energy.

At $Y=0$ RG equations \eqref{two RG equations} simplify to a single equation,
\be
\frac{d g_z}{d \ell} = \frac{g_z^2}{2\pi v_F} ,
\label{eq:20}
\ee
whose solution is given by $g_z(\ell) = g_z(0)/[1  - g_z(0) \ell/(2\pi v_F)]$, and $g_c(\ell)$ is described by the same equation.
Thus $\ell_0 = 2\pi v_F/g_z(0) = 6\pi v_F/g$.
The corresponding gap is exponentially small, $\Delta_\sigma \sim (v_F/\alpha) \exp(-6\pi v_F/g)$.

An important clarification is in order here. Reference \onlinecite{garate2010} has showed that quadratic in spin-orbit interaction terms affect the RG flow significantly, via the change of the initial values of the coupling constants, in the limit $b \ll \alpha_R k_F$. Under these conditions the ground state is actually an anisotropic Luttinger liquid \cite{garate2010,chan2017}. This, however, does not affect the conclusion of the flow towards the strong coupling in the optimal case of $b \approx \alpha_R k_F$, on which we are focusing here.

\subsubsection{Bosonized form}
\label{sec:bos}

The physics of the spin gap phase is conveniently discussed with the help of {\em abelian} bosonization, brief description of which is summarized in Appendix \ref{sec: bosonization periodic}.
With this powerful technique the charge Hamiltonian
$\mathcal{H}_\rho=
\mathcal{H}_\rho^0+\mathcal{H}_\mathrm{int,\rho}+\mathcal{H}_\mathrm{flux}$ 
turns into
\bea
\mathcal{H}_\rho&=&\int\!dx\,\frac{1}{2}\left[
\frac{v_\rho}{K_\rho}:\!(\partial_x\phi_\rho)^2\!:
+v_\rho K_\rho:\!(\partial_x\theta_\rho)^2\!:
\right.\nonumber\\
&&\left.{}\qquad\quad
+\frac{2\sqrt{2\pi}\lambda v_F}{L}\partial_x\theta_\rho
\right],
\label{H_rho}
\eea
where $[\phi_\rho(x),\partial_y\theta_\rho(y)]=i\delta(x-y)$, and
\begin{subequations}
\bea
K_\rho&=&\left(1+\frac{2U_0-U_{2k_F}}{\pi v_F}\right)^{-1/2},
\label{K_rho}
\\
v_\rho&=&v_F\left(1+\frac{2U_0-U_{2k_F}}{\pi v_F}\right)^{1/2}.
\label{v_rho}
\eea
\end{subequations}
Note that the relation $v_\rho K_\rho=v_F$ holds, which is a consequence of Galilean invariance and guarantees the $2\pi$-flux periodicity irrespective to the Coulomb
interaction \cite{capponi2000,starykh2000,seidel2005}.

The non-interacting spin Hamiltonian \eqref{H_sigma^0 M} turns into that of a free conjugated pair of bosons, $\phi_\sigma$ and $\theta_\sigma$,
\be
{\cal H}^0_{\rm \sigma} =
\int dx~ \frac{v_F}{2} [(\partial_x \phi_\sigma)^2 + (\partial_x \theta_\sigma)^2] ,
\label{eq:10B}
\ee
where the bosonic fields obey the commutation relation
$[\phi_\sigma(x),\partial_y\theta_\sigma(y)]=i\delta(x-y)$.
Bosonized form of the interaction \eqref{eq:a3} is obtained as
\bea
{\cal H}_{\rm int, \sigma} &=&
\int dx \Big\{
 -\frac{g_z}{8\pi} [(\partial_x \phi_\sigma)^2 - (\partial_x \theta_\sigma)^2] 
\nonumber\\
&&{}\qquad\quad
- \frac{g_c}{(2\pi \alpha)^2} \cos(\sqrt{8\pi}\theta_\sigma)
\Big\} ,
\label{eq:18}
\eea
where $\alpha$ is a short-distance cutoff.
From \eqref{eq:10B} and \eqref{eq:18} we obtain
\bea
{\cal H}_\sigma &=& \int dx
\Big[\frac{v_\sigma}{2K_\sigma} (\partial_x \phi_\sigma)^2
 + \frac{v_\sigma K_\sigma}{2} (\partial_x \theta_\sigma)^2\nonumber\\
 &&{}\qquad\quad
 - \frac{g_c}{(2\pi \alpha)^2} \cos(\sqrt{8\pi}\theta_\sigma)\Big],
\label{eq:21}
\eea
where we introduced dimensionless Luttinger parameter $K_\sigma$ and renormalized spin velocity $v_\sigma$,
\begin{subequations}
\bea
K_\sigma &=& \sqrt{\frac{1 + g_z/4\pi v_F}{1 - g_z/4\pi v_F}}, 
\label{K_sigma}
\\
v_\sigma &=& v_F \sqrt{1 - (g_z/4\pi v_F)^2},
\label{v_sigma}
\eea
\end{subequations}
and we set
$\eta_{R\uparrow}\eta_{R\downarrow}\eta_{L\uparrow}\eta_{L\downarrow}=1$.
The non-linear cosine term $\cos(\sqrt{8\pi}\theta_\sigma)$ in \eqref{eq:21} describes inter-subband pair-tunneling processes, the Cooper scattering in Fig.~\ref{fig:cooper}, and 
is responsible for the opening of the spin gap.
We note in passing that the above analysis can be easily extended to the case of non-orthogonal spin-orbit and Zeeman field directions \cite{suhas2008,garate2010,handbook2010}. 
In this case $M^+_R M^+_L$ in \eqref{Hint,sigma with oscillations} [or, equivalently, $\cos(\sqrt{8\pi}\theta_\sigma)$ term in \eqref{eq:21}] too acquire oscillating factors and therefore ``average out" 
of the Hamiltonian. Physically, this corresponds to momentum-nonconserving pair-tunneling between the two subbands \cite{suhas2008} and results in the restoration of the critical
Luttinger liquid ground state.

We see that the growth of $g_c$ under RG flow leads to the growth of $K_\sigma$. According to the standard Tomonaga-Luttinger liquid phenomenology \cite{giamarchi2003,gogolin2004}, this indicates 
the development of attractive interactions in the spin sector and associated superconducting-like behavior of various physical observables.
A large positive value of $g_c$ implies the development of the correlated state with $\cos(\sqrt{8\pi}\theta_\sigma) = +1$. This state is two-fold degenerate, with $\sqrt{2\pi}\theta^{(1)}_\sigma = 0$
and $\sqrt{2\pi}\theta^{(2)}_\sigma = \pi$ (mod $2\pi$) describing two equivalent spin states.

The physical meaning of the obtained spin correlated state can be inferred from the behavior of spin density, as was done previously in \cite{suhas2008,garate2010}, and corresponds to the Ising-type
algebraic spin density wave (SDW) order.
Specifically, we are interested in the $2k_F$ component of the spin density
$S^a_{2k_F}(x) = N^a(x) e^{-i 2k_F x} +  {\rm h.c.}$,
where
\be
N^a(x)=\frac{1}{2}\Psi^\dagger_R(x)\sigma^a\Psi^{}_L(x),
\quad
a=x,y,z.
\label{x1}
\ee
With the help of \eqref{rot-fermions} and \eqref{shifting fermi points}, 
$N^a$ reduces to the form
\bea
\label{eq:a44}
N^x &=& 
\cos\beta \widetilde{N}^x+i\sin\beta\cos(t_\varphi x)\widetilde{N}^0
\nonumber\\&&{}
+\sin\beta\sin(t_\varphi x)\widetilde{N}^z,\\
N^y&=&\widetilde{N}^y,\\
N^z &=& \cos(t_\varphi x)\widetilde{N}^z - i \sin(t_\varphi x) \widetilde{N}^0,
\eea
where we have defined
$\widetilde{N}^{a=0,x,y,z} =  \frac{1}{2}\psi^{\dagger}_R \sigma^a \psi_L$, and $\sigma^0$ denotes the identity matrix.
Using the standard bosonization (described in Appendix \ref{sec: bosonization periodic}), we obtain
\begin{subequations}
\bea
\widetilde{N}^0 &=&
\frac{\eta_{R\uparrow}\eta_{L\uparrow}}{2\pi\alpha}
e^{-i\sqrt{2\pi}\phi_\rho-2\pi ix/L} \cos(\sqrt{2\pi}\phi_\sigma), \label{eq:a45.1}\\
\widetilde{N}^x &=&
\frac{i \eta_{R\uparrow} \eta_{L\downarrow}}{2\pi\alpha}
e^{-i\sqrt{2\pi}\phi_\rho-2\pi ix/L} \sin(\sqrt{2\pi}\theta_\sigma), \label{eq:a45.2}\\
\widetilde{N}^y &=&
\frac{-i \eta_{R\uparrow} \eta_{L\downarrow}}{2\pi\alpha}
e^{-i\sqrt{2\pi}\phi_\rho-2\pi ix/L} \cos(\sqrt{2\pi}\theta_\sigma), \label{eq:a45.3}\\
\widetilde{N}^z &=&
\frac{-i\eta_{R\uparrow}\eta_{L\uparrow}}{2\pi\alpha}
e^{-i\sqrt{2\pi}\phi_\rho-2\pi ix/L} \sin(\sqrt{2\pi}\phi_\sigma).
\label{eq:a45.4}
\eea
\label{eq:a45.5}
\end{subequations}
It is now easy to observe that in the ground state of the sine-Gordon Hamiltonian \eqref{eq:21} only the spin part of $\widetilde{N}^y$ acquires a nonvanishing expectation value -- the spin density 
`wants' to line up along the $\hat{y}$-axis, which is the spin-orbit axis; see \eqref{eq:2}. Therefore we can write, choosing the gauge
$\eta_{R\uparrow} \eta_{L\downarrow} = i$ \cite{suhas2008},
\be
\begin{pmatrix}
S^x \\ S^y \\ S^z
\end{pmatrix}_{\!\! 2k_F} \propto
 \cos\!\left[\sqrt{2\pi} \phi_\rho(x) + 2k_F x+\frac{2\pi x}{L}\right]\! 
\begin{pmatrix}
0 \\ \pm 1 \\ 0
\end{pmatrix},
\label{eq:22}
\ee
up to the ``short-ranged'' corrections involving field $\phi_\sigma$, correlation functions of which decay exponentially on the scale $v_\sigma/\Delta_\sigma$.
The $\pm 1$ part of the above equation corresponds to the choice of degenerate ground states $\theta^{(1,2)}_\sigma$.
Gapless charge fluctuations, however, prevent the true symmetry breaking
from happening.
Equations \eqref{eq:a44}, \eqref{eq:a45.5} and \eqref{eq:22} show that spin correlations in the obtained SDW state are highly anisotropic in spin space and their spatial decay is controlled by the gapless charge sector of the wire.

It is also useful to consider the $2k_F$ component of the charge density,
$\rho_{2k_F}^{}(x) = \rho(x) e^{-i 2k_F x} +  {\rm h.c.}$, where
\be
\rho = \Psi^\dagger_R \Psi_L
= \psi^\dagger_R e^{-it_\varphi x\sigma^z/2}e^{i\beta\sigma^x}
e^{-it_\varphi x\sigma^z/2} \psi_L .
\label{eq:a46}
\ee
We find
\be
\rho =
2\cos\beta \cos(t_\varphi x) \widetilde{N}^0
- 2 i \cos\beta \sin(t_\varphi x) \widetilde{N}^z  + 2i \sin\beta \widetilde{N}^x.
\label{eq:a47}
\ee
We see that $\rho_{2k_F}(x)$ is nullified by the SDW ground state. This means that {\em weak} scalar impurity, potential of which couples to $\rho_{2k_F}$, renormalizes to zero -- electron backscattering is suppressed \cite{suhas2008}.

Finally, it is interesting to note that $\mathcal{H}_\sigma$ \eqref{eq:21} at $K_\sigma=2$ is just a bosonized
Hamiltonian of a one-dimensional $p$-wave superconductor, which is known to be a topological superconductor of class D \cite{Kitaev2001,schnyder2008B} having a zero-energy Majorana mode at each end.
Hence we can anticipate that our quantum wire model may also host a localized Majorana-like zero mode at the end of the wire, even though no superconducting order is present in the ground state. This is indeed the case as shown in Sec.~III.

\subsubsection{Ground state of a finite ring}
\label{sec:ring}

From the RG analysis explained above, we have found that the ground state
of the spin gap phase has Ising-type SDW quasi-long range order. 
Here, however, we show that the ground state of the wire is {\em unique} under the ring geometry.

To this end, we need to pay close attention to zero-modes in the low-energy Hamiltonian $H_\rho+H_\sigma$ \cite{loss1992,seidel2005}.
As shown in Appendix \ref{sec: bosonization periodic}, the zero-modes obey the selection rules \eqref{identity 1} and \eqref{identity 2}.
We restrict ourselves to the even particle number parity case when
\be
(-1)^{N_\rho}=(-1)^{J_\rho}=(-1)^{N_\sigma}=(-1)^{J_\sigma}=1,
\label{eq:particle-even}
\ee
and reproduce \eqref{identity 2} here for completeness
\be
(-1)^{\frac{1}{2}(N_\rho+J_\rho)}=(-1)^{\frac{1}{2}(N_\sigma+J_\sigma)}.
\label{identity2b}
\ee
Substituting \eqref{chiral phi} into \eqref{H_rho} and keeping only the zero-mode terms, we find
\be
H_\rho^0=\frac{\pi}{4L}\!\left(
\frac{v_\rho}{K_\rho}N_\rho^2+v_FN_\rho
+v_F J_\rho^2 -4\lambda v_F J_\rho
\right),
\label{eq: H_rho^0}
\ee
which is minimized when $J_\rho=2\lambda$.
We note that $\lambda$ is related to the magnetization; see Eq.~\eqref{flux vs M}.

Similarly, the zero-mode part of the spin Hamiltonian $H_\sigma$ is
\bea
H_\sigma^0&=&
\frac{\pi v_\sigma}{4L}\!
\left(\frac{1}{K_\sigma}N_\sigma^2+K_\sigma J_\sigma^2\right) 
\nonumber\\
&&{}\!
-\frac{g_c\gamma}{(2\pi\alpha)^2}\int^L_0dx\,
\cos(\sqrt{8\pi}\theta_\sigma^0-2\pi J_\sigma x/L),
\label{eq: H_sigma^0}
\eea
where $\theta_\sigma^0$ is defined by
\be
\theta_\sigma^0=\frac{1}{\sqrt2}
(\phi_{L\uparrow}^0-\phi^0_{R\uparrow}
-\phi^0_{L\downarrow}+\phi^0_{R\downarrow}),
\ee
and the renormalization factor $\gamma$ from finite-frequency modes is
\be
\gamma=\left(\frac{2\pi\alpha}{L}\right)^{2/K_\sigma}.
\ee
Assuming that $g_c$ is renormalized to strong coupling,
we find that $H_\sigma^0$
is minimized when
$(J_\sigma,e^{i\sqrt{2\pi}\theta_\sigma^0})=(0,1)$ or $(0,-1)$.
It follows from the commutation relation
\be
[\theta_\sigma^0,N_\sigma]=i\sqrt{\frac{2}{\pi}}
\ee
that
\be
e^{i\sqrt{8\pi}\theta_\sigma^0}N_\sigma e^{-i\sqrt{8\pi}\theta_\sigma^0}
=N_\sigma-4,
\quad
\{e^{i\sqrt{2\pi}\theta_\sigma^0}, e^{i\pi N_\sigma/2}\}=0.
\label{theta_sigma^0,N_sigma}
\ee
We see that $N_\sigma$ is not conserved but the {\em parity} $(-1)^{N_\sigma/2}$ is conserved.

Let us introduce eigenstates of $e^{i\sqrt{2\pi}\theta_\sigma^0}$:
\be
e^{i\sqrt{2\pi}\theta_\sigma^0}|a\rangle = |a\rangle,
\quad
e^{i\sqrt{2\pi}\theta_\sigma^0}|b\rangle = - \, |b\rangle.
\ee
Since the two states $|a\rangle$ and $|b\rangle$ minimize the potential
$-g_c\cos(\sqrt{8\pi}\theta_\sigma^0)$, they are candidates for
ground states of $H_\sigma^0$.
However, they are not eigenstates of a parity operator $(-1)^{N_\sigma/2}$.

Let us define
\be
|+\rangle=\frac{1}{\sqrt2}(|a\rangle + |b\rangle),
\quad
|-\rangle=\frac{1}{\sqrt2}(|a\rangle - |b\rangle).
\ee
We find from \eqref{theta_sigma^0,N_sigma} that $|+\rangle$ and
$|-\rangle$ are, respectively, even and odd parity state,
\be
e^{i\pi N_\sigma/2}|+\rangle=|+\rangle,
\quad
e^{i\pi N_\sigma/2}|-\rangle=-|-\rangle.
\ee
The discussion above follows that in \cite{fidkowski2011} closely.
It follows from \eqref{identity2b} with $J_\sigma=0$ that
\be
(-1)^{\frac{1}{2}(N_\rho+J_\rho)}=\left\{\begin{array}{ll}
+1 & \mbox{for~}|+\rangle, \\
-1 & \mbox{for~}|-\rangle. \\
\end{array}\right.
\ee

We are now ready to see that the ground state of $H_\rho^0+H_\sigma^0$ is unique in the ring geometry.
Since we seek the ground state for a fixed number of electrons, we can set $N_\rho=0$.
\begin{itemize}
\item
Suppose $(-1)^{\frac{1}{2}(N_\rho+J_\rho)}=+1$. With $N_\rho=0$ we have that $J_\rho/2 = {\rm even}$ and can set $J_\rho=0$ to minimize the charge Hamiltonian 
\eqref{eq: H_rho^0}.
By  \eqref{identity2b} we have $(-1)^{\frac{1}{2}(N_\sigma+J_\sigma)} =+1$ too.
Now, \eqref{eq: H_sigma^0} (and also \eqref{eq:21}) is minimized by $J_\sigma=0$ when the field configuration in the argument of the cosine in \eqref{eq: H_sigma^0} and \eqref{eq:21} is uniform, i.e., does not have {\em kinks}. This is easiest to see by thinking of the full field $\theta_\sigma$ in \eqref{eq:21} 
and following its definition in Appendix \ref{sec: bosonization periodic}, see \eqref{x19}.
Then we find $\theta_\sigma(0) = \theta_\sigma(L)$ and the kink-free configuration of $\theta_\sigma$ satisfies this.
With $J_\sigma=0$ we have $(-1)^{N_\sigma/2} = (-1)^M =+1$ and hence the ground state is the state $|+\rangle$. 
Note also that $(-1)^M=1$ means $\lambda=0$, see \eqref{flux vs M}, and therefore the choice of $J_\rho = 0$ indeed corresponds to the energy minimum.

Let us now ask what is the lowest energy for the state $|-\rangle$?
In this state $(-1)^{N_\sigma/2} =-1$ but then our initial assumption $(-1)^{\frac{1}{2}(N_\rho+J_\rho)}=+1$ and \eqref{identity2b} require that $(-1)^{\frac{1}{2}(N_\sigma+J_\sigma)} =+1$. This is only possible if $J_\sigma = \pm 2$. (More generally, $J_\sigma = \pm 2 + 4n$, but this will lead to a multi-kink spin sector configuration with yet higher energy.) But then the field $\theta_\sigma$ must obey 
$\theta_\sigma(L) = \theta_\sigma(0) - \sqrt{\pi/2} J_\sigma$ so that it experiences discontinuity (kink) at $x=0$ (which is the same as $x=L$ in the ring geometry).
This boundary condition forces $\theta_\sigma$ to have another kink somewhere on the ring, at some $0 < x_k < L$. 
The lowest energy of the state with such a 2-kink configuration (one at $x_k$ and another at $x=0=L$) is {\em higher} than that of the kink-free configuration. 
Calculating this energy difference is not easy but the relevance of the cosine potential in \eqref{eq:21} means that it is of the order $g_c/\xi$, where $\xi = \Delta/v_\sigma$ is the correlation length of the correlated SDW state. The energy difference remains finite in the limit $L\to\infty$.

We therefore see that in the case of $(-1)^{\frac{1}{2}(N_\rho+J_\rho)}=+1$  the ground state of the wire is given by $|+\rangle$, i.e., the state with the positive magnetization parity. 
The state with negative magnetization parity $|-\rangle$ has much higher energy. 

\item
Next consider $(-1)^{\frac{1}{2}(N_\rho+J_\rho)}=-1$, which for $N_\rho=0$ means $J_\rho/2 = {\rm odd}$. Now the identity
 \eqref{identity2b} requires $(-1)^{\frac{1}{2}(N_\sigma+J_\sigma)} =-1$. Therefore the kink-free configuration of the spin sector, one with $J_\sigma=0$, requires
 $(-1)^{N_\sigma/2} = (-1)^M=-1$. By \eqref{flux vs M} this means that $\lambda=1$ and hence the charge sector energy is minimized by $J_\rho=2$.
The odd-parity state $|-\rangle$ is the lowest-energy state.

The other, positive magnetization parity state $|+\rangle$ must have finite spin current $J_\sigma = \pm 2$ which therefore forces the spin sector into a 2-kink configuration and results in the higher energy for it.
\end{itemize}

We note that the energy difference between the lowest-energy state $|+\rangle$ under $(-1)^{\frac{1}{2}(N_\rho+J_\rho)}=+1$ and the lowest-energy state $|-\rangle$ under $(-1)^{\frac{1}{2}(N_\rho+J_\rho)}=-1$ is of order $1/L$ due to the charge Hamiltonian \eqref{eq: H_rho^0}.

The presented arguments establish that  the ground state of the wire in the correlated SDW state is unique in the {\em ring geometry}. It is worth noting that the gapless charge sector has played an important role in this conclusion, via the ``super-selection" rules \eqref{eq:particle-even} and especially \eqref{identity2b}. We'll see below that this is not the case in the case of the open wire, i.e., the wire with two open ends.

\section{Finite wire with open boundaries}
\label{sec:obc}

Now we turn to the case of our main interest, i,e., a finite wire with open boundaries at $x=0$ and $x=L$, where $\Psi_s(x=0)=0 = \Psi_s(x=L)$.
Equation \eqref{eq:4} shows that OBC for the original fermions means $\Psi_{R s}(0) = - \Psi_{L s}(0)$ and $\Psi_{R s}(L) = - \Psi_{L s}(L)$. 
The last relation follows from $e^{ik_FL} = 1$, as explained below \eqref{eq:4}.

In order for the {\em rotated} fermions to obey simple boundary conditions which do not mix components with different spin indices $s$, it proves very convenient to change
the direction of the external magnetic field to be along the $\hat{x}$ axis, $\vec{B} = B \hat{x}$, and not along the $\hat{z}$ axis as written in \eqref{eq:3}. This choice does not change the physics of the problem because the magnetic field and spin-orbit interaction remain orthogonal to each other and, therefore, the correlated SDW phase is preserved.
Detailed arguments in Appendix \ref{App:openBC} show how this chiral rotation about the $\hat{z}$ axis  is done and, following the steps described there, one finds that the rotated fermions $\psi_{L/R,s}(x)$ introduced in \eqref{eq:a12} satisfy the boundary condition \eqref{eq:ca19}.
In terms of $s$-components it is just
\be
\psi_{L, s}(x_o) =  - e^{i s\beta}  \psi_{R, s}(x_o),
\label{eq:a19B}
\ee
where up/down spin projection $s= \uparrow = +1, s = \downarrow = -1$ in the {\em rotated} basis and $x_o=0, L$ denotes the two open ends of the wire.

After the chiral rotation \eqref{Rotation}, the total magnetic field $h= \sqrt{b^2 + (2\alpha_R k_F)^2}$ experienced by electrons is pointing along the $\hat{x}$-axis.
Subsequent manipulations (summarized as steps 1-3 in Appendix \ref{App:openBC}) are needed in order to absorb $h$ into the redefined Fermi momenta. 
Therefore up- and down-pointing spins $s = \uparrow, \downarrow$ in \eqref{eq:a19B} and elsewhere in this Section
actually represent spins pointing along the positive and negative $\hat{x}$-axis in the rotated frame.

The resulting Hamiltonian is split into charge Hamiltonian and spin Hamiltonian, and is written in terms of charge currents $J_{R/L}(x)$ \eqref{eq:charge-current1}, which are not affected by the performed rotations, and spin currents $K_{R/L}^a(x)$, which are related to $\psi_{R/L}$ by \eqref{eq:a20}. Both types of currents are expressed in terms of the rotated fermions $\psi_{R/L}(x)$. 

We are now ready to write down the spin Hamiltonian of the wire of finite length $L$ with open boundaries at $x_o = 0, L$. 
It is formulated in terms of right-moving current $\vec{K}_R(x)$ and reads [see \eqref{eq:a24} and \eqref{eq:a26}]
\bea
{\cal H}_{\rm \sigma} &=& {\cal H}^0_{\rm \sigma} + {\cal H}_{\rm int, \sigma},\nonumber\\
\label{eq:a199}
{\cal H}^0_{\rm \sigma} &=& 2\pi v_F \int_{-L}^L dx \, [K^z_R(x)]^2, \\
{\cal H}_{\rm int, \sigma} &=& -\frac{1}{2} \int_{-L}^L \!dx \, 
\Big(g_c K^z_R(x) K^z_R(-x) \nonumber\\
&&\qquad\quad
+ \frac{g_x + g_c}{4}[e^{-i 2\beta} K^+_R(x) K^+_R(-x) + {\rm h.c.}]\Big).
\nonumber\\&&
\label{eq:a200}
\eea

Our next task is to bosonize ${\cal H}_{\rm \sigma}$. 
The first line of \eqref{eq:a200} represents quadratic correction to \eqref{eq:a199}. Using \eqref{eq:a9} we collect quadratic boson terms of ${\cal H}_{\rm \sigma}$,
\bea
{\cal H}^{(2)}_{\rm \sigma} &=& \!
\int_{-L}^L \!dx
\Big\{ v_F [\partial_x \Phi_{R \sigma}(x)]^2
 + \frac{g_c}{4\pi} \partial_x \Phi_{R \sigma}(x) \partial_x \Phi_{R \sigma}(-x)\Big\}
\nonumber\\
&&{}
+\frac{\pi v_\sigma M^2}{L K_\sigma}.
\label{eq:a27}
\eea
This part can be diagonalized with the help of another chiral boson field $\tilde{\Phi}$ (see Ch.~27 of \cite{gogolin2004})
\be
\label{eq:31}
\Phi_{R \sigma}(x) 
= \frac{\sqrt{K_\sigma}}{2} [ \tilde{\Phi}(x) - \tilde{\Phi}(-x)]
 + \frac{1}{2 \sqrt{K_\sigma}} [ \tilde{\Phi}(x) + \tilde{\Phi}(-x)] , 
\ee
where Luttinger parameter $K_\sigma$ is introduced in \eqref{K_sigma}.
Observe that under this transformation, 
\be
\Phi_{R \sigma}(x) + \Phi_{R \sigma}(-x) = \frac{1}{\sqrt{K_\sigma}} [ \tilde{\Phi}(x) + \tilde{\Phi}(-x)] .
\label{eq:o10}
\ee

The nonlinear operator in the second line of \eqref{eq:a200} is found, with the help of Baker-Hausdorff identity $e^A e^B = e^{A + B} e^{[A,B]/2}$, to be
\bea
e^{-i 2\beta} K^+_R(x) K^+_R(-x)
&=& \frac{(F_\uparrow^\dagger F_\downarrow)^2}{(2\pi \alpha)^2}
 e^{-i\sqrt{8\pi}[\Phi_{R \sigma}(x) + \Phi_{R \sigma}(-x)]} \nonumber\\
&&\times e^{-2\pi ix/L}  e^{-4\pi \Upsilon(x)}
\nonumber\\
&=&\!
-\frac{(F_\uparrow^\dagger F_\downarrow)^2}{(2\pi \alpha)^2}
 e^{-i\sqrt{\frac{8\pi}{K_\sigma}}[\tilde{\Phi}(x) + \tilde{\Phi}(-x)]} .
\nonumber\\&&
\label{eq:o9}
\eea
Here
\bea
\Upsilon(x) &=& [\Phi_{R \sigma}(x), \Phi_{R \sigma}(-x)]
= \frac{1}{4\pi} \ln\!\left(
\frac{e^{\pi \alpha/L} - e^{-i 2\pi x/L}}{e^{\pi \alpha/L} - e^{i 2\pi x/L}}
\right) \nonumber\\
&\to& \frac{i}{2\pi} \tan^{-1}\!\left[\cot\!\left(\frac{\pi x}{L}\right)\right],
\label{eq:a28}
\eea
which is obtained from \eqref{eq:a7}. The last line represents the limit $\alpha/L \to 0$. This leads to $e^{-i \frac{2\pi x}{L}}  e^{-4\pi \Upsilon(x)} = -1$ in \eqref{eq:o9}.

Putting everything together, we find 
\begin{subequations}
\be
{\widetilde{\cal H}}_\sigma 
= \frac{\pi v_\sigma M^2}{L K_\sigma}
+\int_{-L}^L \!dx\,\widetilde{H}_\sigma,
\label{eq:31A}
\ee
where
\bea
\widetilde{H}_\sigma
&=&
\frac{g_x + g_c}{8 (2\pi \alpha)^2}\!\left[
(F_\uparrow^\dagger F_\downarrow)^2 
e^{-i \sqrt{\frac{8\pi}{K_\sigma}} [\tilde{\Phi}(x) +  \tilde{\Phi}(-x)] } + {\rm h.c.}\right]
\nonumber\\
&&{}
+ v_\sigma [\partial_x \tilde{\Phi}(x)]^2 .
\label{tilde H_sigma}
\eea
\end{subequations}
Equation \eqref{eq:31A} is chiral version of \eqref{eq:21},
with $\theta_\sigma$ rescaled by $\sqrt{K_\sigma}$. 

It is worth noting here that in the open wire geometry the charge $J_\rho$ and the spin $J_\sigma$ currents are necessarily absent \cite{fabrizio1995}, and as a result 
the zero-mode part of \eqref{eq:31A} consists of a single term $\propto M^2/L$. This also means that the ``super-selection" rule \eqref{eq:particle-even} reduces to
$(-1)^{N_\rho}=(-1)^{N_\sigma}$ while \eqref{identity2b}, with $J_\rho=J_\sigma=0$, becomes its natural consequence. Altogether, this means that global zero-mode constraints \eqref{eq:particle-even} and \eqref{identity2b}, which played a crucial role in the ring geometry in Sec.~\ref{sec:ring}, largely lose their importance in the open wire geometry.

We now observe that at a special value $K_\sigma = 2$, which defines the Luther-Emery point \cite{luther1974,delft1998}, the cosine term in \eqref{tilde H_sigma} is proportional to the product 
of $e^{-i\sqrt{4\pi} \tilde{\Phi}(x)}$ and 
$e^{-i\sqrt{4\pi} \tilde{\Phi}(-x)}$, suggesting, by comparison with \eqref{eq:a7}, that it can be written as a bilinear form of fermion-like operators.
Therefore, at $K_\sigma = 2$ a re-fermionization is possible. To that end, we introduce the new spinless fermion operator via
\be
f(x) = \frac{1}{\sqrt{2\pi \alpha}} {\cal F}
 e^{i \pi x M/L} e^{i \sqrt{4\pi} \tilde{\Phi}(x)} ,
\label{eq:32}
\ee
where ${\cal F}\equiv F_\downarrow^\dagger F_\uparrow^{}$ is a new Klein factor. 
Observe that it satisfies all requirements of being the $M$-changing operator,
\be
[M, {\cal F}] = -{\cal F}, \quad
[M, {\cal F}^\dagger] = {\cal F}^\dagger, \quad
{\cal F}^\dagger {\cal F} = {\cal F} {\cal F}^\dagger = 1.
\label{eq:o11}
\ee
The exponential in \eqref{tilde H_sigma} can now be re-written, at $K_\sigma = 2$, as 
\bea
({\cal F}^\dagger)^2 e^{-i\sqrt{4\pi}[\tilde{\Phi}(x) + \tilde{\Phi}(-x)]} \!
&=&
e^{-i\sqrt{4\pi}\tilde{\Phi}(x)} {\cal F}^\dagger e^{-i\sqrt{4\pi}\tilde{\Phi}(-x)}
\nonumber\\
&&\times{\cal F}^\dagger  e^{2\pi \Upsilon(x)}
\nonumber\\
&=&
e^{-i\sqrt{4\pi}\tilde{\Phi}(x)} {\cal F}^\dagger e^{- i \frac{\pi x M}{L}}
\nonumber\\
&&\!\times
e^{i \frac{\pi x M}{L}} e^{-i\sqrt{4\pi}\tilde{\Phi}(-x)}
{\cal F}^\dagger  e^{2\pi \Upsilon(x)}
\nonumber\\
& = &
2\pi \alpha f^\dagger(x) f^\dagger(-x) \, e^{i \frac{\pi x}{L}} e^{2\pi \Upsilon(x)}.
\nonumber\\&&
\label{eq:a30}
\eea  
To obtain the last equality above, we used 
${\cal F}^\dagger e^{- i \pi x M/L} = e^{- i \pi x M/L} {\cal F}^\dagger e^{i \pi x/L}$.
Observe that $[\tilde{\Phi}(x), \tilde{\Phi}(-x)] = \Upsilon(x)$ and that for $|x| \gg \alpha$ \eqref{eq:a28} gives
\be
e^{2\pi\Upsilon(x)} e^{i \pi x/L} =  i s(x) ,
\label{eq:o14}
\ee
where $s(x) = \sgn[\sin \pi x/L]$ is a $2L$-periodic sign-function
\bea
s(x) = 
\left\{
\begin{array}{rl}
1, & x\in (0,L),\\
-1, & x\in (-L,0),\\
0, & x= 0, \pm L .\\
\end{array}\right.
\label{eq:a29}
\eea
Keeping small but finite $\alpha$ in \eqref{eq:a28} rounds discontinuities of $s(x)$ in finite intervals of order $\alpha$ around end-points $x=x_o$.
Equations \eqref{eq:o9} and \eqref{eq:a30} show that at $K_\sigma=2$
\be
e^{-i 2\beta} K^+_R(x) K^+_R(-x) = \frac{-i s(x)}{2\pi \alpha} f^\dagger(x) f^\dagger(-x).
\label{eq:a31}
\ee
The final ingredient is the kinetic energy which we, following \cite{delft1998} and using formalism developed in Appendix \ref{app:Bos-OBC}, find to be 
\bea \!\!\!
\int_{-L}^L \!dx \, f^\dagger(x) (-i v_\sigma \partial_x) f(x) 
&=& \frac{\pi v_\sigma}{2L} M(M+1) \nonumber\\
&&\!\!\!{}
+ \!\int_{-L}^L \! dx \, v_\sigma [\partial_x \tilde{\Phi}(x)]^2 . ~
\label{eq:a31 2nd}
\eea
Hence at $K_\sigma=2$ the spin Hamiltonian \eqref{eq:31A} can be written in terms of new fermion operators \eqref{eq:32} as
\bea
{\widetilde{\cal H}}_\sigma &=&
\!\int_{-L}^L \!dx \left[ f^\dagger(x) (-i v_\sigma \partial_x) f(x) 
+ \frac{i\Delta}{2} s(x)  f(x) f(-x)
\right.
\nonumber\\
&&\left.\qquad\quad
 + \frac{i\Delta}{2} s(x)  f^\dagger(x) f^\dagger(-x) \right]
 - \frac{\pi v_\sigma M}{2L},
\label{eq:34}
\eea
where the spin gap is given by $\Delta = (g_x+g_c)/(8\pi \alpha)$.
Equation \eqref{eq:34} describes a one-dimensional $p$-wave superconductor with the pairing potential changing sign at $x = x_o =0, \pm L$ \cite{loss2020}, i.e., at the open boundaries
of the wire in our original problem. The sign function in \eqref{eq:34} is required because fermions anti-commute,  $\{f(x),f(-x)\}=0$.
The kink in $\Delta$ is of the profound importance to the low-energy excitations.
We show below that it induces a zero-energy self-conjugate state, the Majorana mode, which
is exponentially localized near the boundary. 

Note that although the Hamiltonian \eqref{eq:34} does not conserve $M$ due to the presence of the Klein factors ${\cal F}$ in $f(x)$, it does conserve
$e^{i 2\pi M} = e^{i \pi (N_\uparrow - N_\downarrow)}$, which follows from $[e^{i 2\pi M}, {\cal F}]=0$.

Moreover, \eqref{eq:34} also conserves the {\em magnetization parity} $e^{i \pi M}$, 
\be
e^{i \pi M} = e^{i \frac{\pi}{2} (N_\uparrow - N_\downarrow)} .
\label{eq:magparity}
\ee
 This is because \eqref{eq:34} contains {\em squares} of Klein factors ${\cal F}$ and ${\cal F}^\dagger$.
As a result, we have to look on $e^{i \pi M} {\cal F}^2$. However, the commutation relation \eqref{eq:o11} implies that $[M, {\cal F}^2] = - 2 {\cal F}^2$ and therefore 
$e^{i \pi M} {\cal F}^2 = {\cal F}^2 e^{i \pi M} e^{-i 2\pi} =  {\cal F}^2 e^{i \pi M}$. That is, $[e^{i \pi M}, {\widetilde{\cal H}}_\sigma]=0$, the magnetization parity \eqref{eq:magparity} is conserved by the Hamiltonian \eqref{eq:34}.

The charge sector of the open wire is described in Appendix \ref{app:obc-charge}.

\section{Solution of the refermionized Hamiltonian}
\label{sec:refermion}

One-dimensional superconductor \eqref{eq:34} is solved by the Bogolyubov transformation \cite{DeGennes}
\be
f(x) = \sum_{n\ge0} \left[\gamma_n u_n(x) + \gamma_n^\dagger v^*_n(-x) \right],
\label{eq:35}
\ee
which diagonalizes \eqref{eq:34} into the form
\be
{\widetilde{\cal H}}_\sigma = E_{\rm g.s.} + \sum_n \epsilon_n \gamma_n^\dagger \gamma_n .
\label{eq:36}
\ee
Here $\epsilon_n \geq 0$ are the excitation energy, and $\gamma_n$ are Fermi operators satisfying $\{\gamma_n, \gamma_m^\dagger\} = \delta_{n,m}$ and 
$\{\gamma_n, \gamma_m\} = 0$.
Functions $u(x), v(x)$ are found with the help of the equation of motion
\bea
i\partial_t f(x) &=& [f(x), {\widetilde{\cal H}}_\sigma]
\nonumber\\
&=&
-\frac{\pi v_\sigma}{2L} f(x) - i v_\sigma \partial_x f(x)
- i \Delta \sgn(x) f^\dagger(-x)
\nonumber\\&&
\label{eq:37}
\eea
by expressing both sides of the last equality in terms of fermion operators $\gamma_n, \gamma_n^\dagger$ with the help of \eqref{eq:36} and \eqref{eq:37}. 
This leads to the Bogolyubov - de Gennes equation
\be
\begin{pmatrix} 
-i v_\sigma \partial_x -\frac{\pi v_\sigma}{2L} & i \Delta s(x) \\
 -i \Delta s(x) & i v_\sigma \partial_x  +\frac{\pi v_\sigma}{2L}
\end{pmatrix}
\begin{pmatrix} 
u(x) \\
v(x)
\end{pmatrix}
= \epsilon \begin{pmatrix} 
u(x) \\
v(x)
\end{pmatrix}.
\label{eq:38}
\ee
What are boundary conditions for $f(x)$ and, as a result, for $u(x)$ and $v^*(-x)$?
Equation \eqref{eq:32} shows that $f(x+2L) = e^{i 2\pi M} f(x)$, 
where we used $2L$-periodicity of $\tilde{\Phi}(x)$ and the commutation relations \eqref{eq:o11}. Correspondingly, the vector 
$(u(x), v^*(-x))^T$ has the same boundary conditions as $f(x)$.
It then follows that
\be
\begin{pmatrix} 
u(L) \\
v(L)
\end{pmatrix}
= e^{i \Theta} 
\begin{pmatrix} 
u(-L) \\
v(-L)
\end{pmatrix} ,
\quad
\Theta = 2\pi M.
\label{eq:o18}
\ee
Note that at $x=0$ the vector $(u(x),v(x))^T$ is continuous.

As noted below \eqref{eq:34}, even though $M$ is not conserved by ${\widetilde{\cal H}}_\sigma$, the exponential $e^{i 2\pi M}$ remain unchanged and is conserved,
because anomalous $f f$ and $f^\dagger f^\dagger$ 
terms in ${\widetilde{\cal H}}_\sigma$ change $M$ by $\pm 2$.
Therefore we can treat $\Theta$ in \eqref{eq:o18} as a real (non-operator) phase, but distinguish the cases of $M = {\rm integer}$ and $M = \mbox{half-integer}$,
\be
\begin{split}
\Theta &= 0 \,\, {\rm for} \,\, M \in \mathbb{Z},
\\
\Theta &= \pi \,\, {\rm for} \,\, M \in \mathbb{Z}+1/2 .
\end{split}
\label{eq:o25}
\ee

Particle-hole symmetry of \eqref{eq:38} ensures that vector $(v^*(-x), u^*(-x))^\mathrm{T}$ describes states with energy $-\epsilon$ and satisfies \eqref{eq:o18}.

Full solution of \eqref{eq:38} 
consists of scattering states $\tilde{f}_\epsilon$ with energies $\epsilon$ above the gap $\Delta$ and a localized in-gap states $f_0, f_L$
with nearly zero energy,
\be
f(x) = f_0(x) + f_L(x) + \int_\Delta^\infty d\epsilon ~\tilde{f}_\epsilon(x) .
\label{eq:43}
\ee

We focus on the localized modes which, for $0 < x < L$, are described by  
\be
\begin{pmatrix} 
u_0(x) \\
v_0(x)
\end{pmatrix} 
= \frac{A_1}{\sqrt{2}} \! \begin{pmatrix} 
1\\
- e^{i \phi}\\
\end{pmatrix}\!
 e^{(- \kappa + i \frac{\pi}{2L})x}
+ \frac{B_1}{\sqrt{2}} \! \begin{pmatrix} 
1\\
e^{-i \phi}\\
\end{pmatrix}\!
 e^{ (\kappa +i \frac{\pi}{2L})x} .
\label{eq:39}
\ee
The corresponding energy is $\epsilon = \sqrt{\Delta^2 - v_\sigma^2 \kappa^2}$, and we introduced $e^{i \phi} = (v_\sigma \kappa + i \epsilon)/\Delta$.
Solution on the negative half of the wire, $-L < x <0$, is given by the similar combination with amplitudes $A_2, B_2$ and $\Delta \to -\Delta$
due to the oddness of the function $s(x)$. Boundary condition \eqref{eq:o18} and continuity of $(u(x),v(x))^T$ at $x=0$ can be written
in the form of $4\times 4$ matrix equation, acting on the vector $(A_1, B_1, A_2, B_2)^T$, with zero right-hand side.
Setting determinant of that matrix to zero produces the relation between $\epsilon$ and $\kappa$,
\be
\tan^2(\phi) \sinh^2(\kappa L) = \cos^2(\Theta/2),
\label{eq:o19}
\ee
which can be used to express everything in terms of $\kappa$ as
\bea
\Delta &=& v_\sigma \kappa \sqrt{1+ \frac{\cos^2\Theta/2}{\sinh^2(\kappa L)}}, \nonumber\\
\epsilon &=&  v_\sigma \kappa \frac{\cos\Theta/2}{\sinh(\kappa L)} \approx 2  \Delta \cos(\Theta/2) e^{-\Delta L/v_\sigma}.
\label{eq:o20}
\eea
We used \eqref{eq:o25} which guaranties that $\cos\Theta/2 \geq 0$.
It is worth pointing out the surprising feature of the vanishing splitting $\epsilon$ between the first excited state, localized at the opposite ends of the wire, 
and the ground state of the wire for the special value of the phase difference $\Theta = \pi$ (mod $2\pi$), when $M \in \mathbb{Z}+1/2$ is half-integer.
The energy splitting $\epsilon$ is maximal when $M \in \mathbb{Z}$,
which corresponds to {\em even} $2 M = N_\uparrow - N_\downarrow$.
Similar oscillatory dependence on the phase difference $\Theta$ has been previously studied in \cite{Vionnet2017} in a different context.

The most important physical message of \eqref{eq:o20} is that the first excited state is exponentially close to the ground state. For $L \gg \xi=v_\sigma/\Delta$ it is 
essentially degenerate with it.

Straightforward algebra leads to 
\bea
A_1 &=& e^{2 \kappa L}(1 - i e^{ - \kappa L + i \Theta/2}) {\cal C},
\quad
B_1 = i(e^{\kappa L + i \Theta/2} + i) {\cal C},\nonumber\\
A_2 &=& B_1^*, 
\quad
B_2=A_1^*.
\label{eq:o21}
\eea
Here ${\cal C}$ is the real normalization constant. To the leading order ${\cal C} = \sqrt{\kappa/2} e^{-2 \kappa L}$.
Therefore, for $0 < x < L$,
\bea
\begin{pmatrix} 
u_0(x) \\
v_0(x)
\end{pmatrix} &\approx &
\frac{\sqrt{\kappa}}{2} \left[\begin{pmatrix} 
~1 \\
-1\\
\end{pmatrix} e^{- \kappa x + i \pi x/2L}
\right.\nonumber\\ && \left.\qquad
+ i e^{i \Theta/2} \begin{pmatrix} 
1 \\
1\\
\end{pmatrix} e^{\kappa (x-L) + i \pi x/2L}\right],
\label{eq:o22}
\eea
where we neglected exponentially small corrections $\propto e^{-\kappa L}$ to the two end-contributions describing exponentially localized in-gap states 
near the left, $x=0$, and the right, $x=L$, ends of the wire. We also used \eqref{eq:o20} to replace $e^{\pm i \phi}$ by $1$ upto exponentially small terms.
For $-L < x < 0$ we similarly find
\bea
\begin{pmatrix} 
u_0(x) \\
v_0(x)
\end{pmatrix} &\approx &
\frac{\sqrt{\kappa}}{2} \left[ -i e^{-i \Theta/2} \begin{pmatrix} 
1 \\
1\\
\end{pmatrix} e^{-\kappa (x+L) + i \pi x/2L}
\right.\nonumber\\ && \left. \qquad
+  \begin{pmatrix} 
~1 \\
-1\\
\end{pmatrix} e^{\kappa x + i \pi x/2L}\right].
\label{eq:o22B}
\eea

Equations \eqref{eq:35},  \eqref{eq:o22} and \eqref{eq:o22B}  finally allow us to express exponentially localized end modes as 
\be
f_0(x) \approx \frac{\sqrt{\kappa}}{2} e^{-\kappa x} (\gamma_0 - \gamma_0^\dagger) \quad{\rm for} ~0< x \ll L,
\label{eq:o23}
\ee
and 
\be
f_L(x) \approx -\frac{\sqrt{\kappa}}{2} e^{-\kappa (L - x)} e^{i \Theta/2} (\gamma_0 + \gamma_0^\dagger) \quad{\rm for} ~L-x \ll L.
\label{eq:o24}
\ee
Note that \eqref{eq:o23} and \eqref{eq:o24} describe self-conjugate Majorana modes,
$f_0 = - f_0^\dagger$ and $f_L^\dagger = e^{-i \Theta} f_L$.
Namely, $f_0$ and $f_L$ are proportional to the independent Majorana modes, $d_1$ and $d_2$ defined via 
\be
\gamma_0 = \frac{1}{\sqrt{2}}(d_1 + i d_2)
\label{eq:a32}
\ee
with
\be
\{d_1,d_2\}=0, \quad (d_1)^2 = (d_2)^2 = \frac{1}{2},
\ee
such that
\be
f_0(0) = i \sqrt{\frac{\kappa}{2}} d_2,
\quad
f_L(L) =  -e^{i \Theta/2} \sqrt{\frac{\kappa}{2}} d_1 .
\label{eq:a33}
\ee

The appearance of the independent Majorana modes at the opposite ends of the quantum wire agrees with the original proposal of Kitaev \cite{Kitaev2001}.
The ground state of the wire is doubly degenerate since 
states $|0\rangle$ and $|1\rangle = \gamma_0^\dagger |0\rangle$ have the same energy \cite{Kitaev2001}, up to exponentially small energy difference 
$\epsilon$ given by \eqref{eq:o20}. 
Here $|0\rangle$ is the vacuum state of $\gamma_n$, $\gamma_n|0\rangle=0$
for $n\ge0$.

Finally, we comment on the applicability of the present analysis away from $K_\sigma=2$.
It is likely from \eqref{K_sigma} that the bare value of the Luttinger parameter $K_\sigma$ is smaller than 2 for not too strong coupling ($g_z/4\pi v_F<3/5$).
However, $K_\sigma$ is renormalized according to the RG equation \eqref{two RG equations} and reaches $K_\sigma=2$ at some length scale, at which we can apply the referemionization.
In this sense the analysis above is applicable to broader range of parameters.
Physically, the two-fold ground-state degeneracy is a direct consequence of the SDW order.

\section{Physical meaning of the Majorana mode}
\label{sec:maj}

\subsection{Spin correlations in the bulk of the wire}
\label{sec:bulk}

Spin excitations of the open wire consist of massive propagating modes $\tilde{f}_\epsilon$, with energy $\epsilon \geq \Delta$, and zero-energy modes $f_{0,L}$ which are exponentially localized on the scale $\xi = \kappa^{-1} = v_\sigma/\Delta$ near $x_o = 0$ and $x_o=L$, correspondingly. We therefore expect that spin correlations inside the open wire, 
for $\xi \ll x \ll L - \xi$, should coincide with those in the ring geometry, see Sec.~\ref{sec:bos}.

To see how this comes about, we start with $K_R^+(x)$, \eqref{x2},
\be
K_R^+(x) \propto e^{-i \sqrt{4\pi} [\tilde{\Phi}(x)-\tilde{\Phi}(-x)]  - i \sqrt{\pi} [\tilde{\Phi}(x) + \tilde{\Phi}(-x)]} 
\label{x6}
\ee
and observe that according to \eqref{eq:31A} the SDW ordered state is characterized by the  ordered, or ``frozen", symmetric combination of spin fields $\tilde{\Phi}(x) +  \tilde{\Phi}(-x)$.
Importantly, the antisymmetric combination $\tilde{\Phi}(x)-\tilde{\Phi}(-x)$ does not commute with the symmetric one. Indeed, simple calculation, similar to one in \eqref{eq:a28}, shows that for $-L < x, y < L$
\bea
&&[\tilde{\Phi}(x) +  \tilde{\Phi}(-x), \tilde{\Phi}(y) - \tilde{\Phi}(-y)] 
= \frac{i}{2}\Big( \sgn(x-y) \nonumber\\
&&~ - \sgn(x+y) + \frac{2y}{L}\Big) .
\label{eq:a52}
\eea
Therefore the ordering (freezing) of the symmetric combination $\tilde{\Phi}(x) +  \tilde{\Phi}(-x)$ makes correlations of operators 
involving the antisymmetric one $\tilde{\Phi}(x) -  \tilde{\Phi}(-x)$ short-ranged, e.g., decaying exponentially with distance. 

Now we turn to the $2k_F$ component of the spin density $N^a$ \eqref{x1} and its rotated version \eqref{x3}, \eqref{x4}. Analysis in Appendix \ref{app:observe} shows 
that $\widetilde{N}^{x,y}$ fields involve the symmetric spin mode $\tilde{\Phi}(x) + \tilde{\Phi}(-x)$ as well as an antisymmetric charge one $\Phi_{R \rho}(x) - \Phi_{R \rho}(-x)$.
In the SDW phase the symmetric spin mode is frozen, but charge excitations remain critical. Moreover, Eq.\ \eqref{xxx5} and discussion around it show that, when the frozen value of the symmetric mode is substituted, the spin part 
of $\widetilde{N}^{x}$ vanishes while that of $\widetilde{N}^y$ approaches a constant value. Correspondingly, correlations of $\widetilde{N}^{y}$ field inside the wire decay algebraically with the exponent which is controlled by the Luttinger constant of the charge mode $K_\rho$, in agreement with expressions \eqref{eq:a45.3} and \eqref{eq:22} for the closed wire case. At the same time, correlations of components $\widetilde{N}^{0,z}$ decay exponentially with the distance, because they involve quantum-disordered antisymmetric spin combination 
$\tilde{\Phi}(x) -  \tilde{\Phi}(-x)$. 

The described correspondence also shows that $\tilde{\Phi}(x) + \tilde{\Phi}(-x)$ plays the role of $\theta_\sigma$, while $\tilde{\Phi}(x) - \tilde{\Phi}(-x)$ is analogous to 
$\phi_\sigma$ in Sec.~\ref{sec:bos}.

\subsection{End-to-end correlations}
\label{sec:end-end}

Correlations between the end regions of the wire, $0 < x < \xi$ and $L - \xi < x < L$, are very different. Observe that for $x\approx 0$ the quantum-disordered combination 
vanishes, $\tilde{\Phi}(x) -  \tilde{\Phi}(-x) \approx 0$. The same is true for $x\approx L$ due to the $2L$-periodicity of the field $\tilde{\Phi}(x)$. At the same time the symmetric combination simplifies to $\tilde{\Phi}(x) + \tilde{\Phi}(-x)\approx 2\tilde{\Phi}(0)$. [Obviously, for $x\approx L$ we have $\tilde{\Phi}(x) + \tilde{\Phi}(-x)\approx 2\tilde{\Phi}(L)$.]
Using \eqref{x2} and the definition of the fermion $f(x)$ \eqref{eq:32}, we observe that 
\be
\begin{split}
K^+_R(0) &= \frac{e^{i\beta}}{\sqrt{2\pi\alpha}} f^\dagger(0), \\
K^+_R(L) &= -\frac{e^{i\beta}}{\sqrt{2\pi\alpha}} e^{-i \pi M} f^\dagger(L).
\end{split}
\label{x7}
\ee
Equation \eqref{eq:a33} shows that fermion operators at the ends of the wire reduce to the Majorana modes $d_{1,2}$, and therefore the same is true for the spin currents $K_R^+$ at the ends of the chain. Note, however, the appearance of the ``string" operator 
$e^{-i \pi M}$ in $K^+_R(L)$ in \eqref{x7}.  This string operator is in fact the {\em magnetization parity}, introduced in Eq. \eqref{eq:magparity}. 
It represents a key integral of motion of the problem and plays a very important role in the subsequent analysis of the open wire.

It is this string operator that makes sure that spin densities at $x \sim 0$ and $x\sim L$ 
actually commute, as they must do (and not anti-commute, as they would if it was absent). 
At this point it is important to observe that $e^{\pm i \pi M}$ and $f(x)$, introduced in \eqref{eq:32}, anticommute for all $x$. 
This is easy to see with the help of identity (C9) of Ref.~\cite{delft1998}
and \eqref{eq:o11}. Hence
\be
\{e^{\pm i \pi M}, f(x)\} = 0 = \{e^{\pm i \pi M}, f^\dagger(x)\}.
\label{eq:a35}
\ee
Next, Eq.~\eqref{eq:35} implies that, for all $n$,
\be
\{e^{i \pi M}, \gamma_n\} = 0 = \{e^{i \pi M}, \gamma^\dagger_n\} .
\label{eq:a36}
\ee
Therefore we can establish an operator identity
\be
e^{i \pi M} = e^{i \Theta/2} \prod_\ell e^{i \pi \gamma^\dagger_\ell \gamma_\ell}
= e^{i \Theta/2} \prod_\ell (1 - 2 \gamma^\dagger_\ell \gamma_\ell),
\label{eq:a37}
\ee
which enforces Eqs.\ \eqref{eq:a36} and \eqref{eq:a35} and also insures that $e^{i 2\pi M} = e^{i \Theta} =  \pm 1$, since 
$(1 - 2 \gamma^\dagger_\ell \gamma_\ell)^2 = 1$ for every $\ell$. Here phase $\Theta$ is the c-number
introduced in \eqref{eq:o18}. Note that \eqref{eq:a37} {\em does not} mean that $M = \Theta/(2\pi) + \sum_\ell \gamma^\dagger_\ell \gamma_\ell$.

At very low energies $\epsilon \ll \Delta$ 
\be
e^{i \pi M} \approx e^{i \Theta/2}(1 - 2\gamma_0^\dagger \gamma_0) = -2 i e^{i \Theta/2} d_1 d_2 .
\label{eq:a38}
\ee
Therefore, while $f^\dagger(L) \sim d_1$, the magnetization parity acting on it changes it into the Majorana mode $d_2$, $e^{-i \pi M} f^\dagger(L) \sim  (d_1 d_2) d_1 = -d_2$.
More accurately, we obtain 
\be
K^+_R(L) = -\frac{e^{i\beta}}{\sqrt{2\pi\alpha}} e^{-i \Theta} f^\dagger(0) = - e^{-i \Theta} K^+_R(0).
\label{eq:a39}
\ee
Therefore spin currents at the opposite ends of the wire are equal, up to a complex pre-factor. 
We see that they commute, $[K^+_R(L), K^+_R(0)]=0$, as they should. Note that without the string operator  $e^{-i \pi M}$ in \eqref{x7} the spin currents in question would 
anti-commute, $\{K^+_R(L), K^+_R(0)\}=0$, just as fermion operators do. The string operator is crucial for obtaining the correct result. 
From the quantum computing point of view, the string operator spoils braiding statistics of the localized end modes \cite{tserkovnyak2011}.

We also observe that $K^+_R(-L) = K^+_R(L)$, thanks to $e^{i 2\pi M} = e^{-i 2\pi M}$ for the integer/half-integer $M$. Last line in \eqref{eq:a9} allows us to 
write the total spin current in terms of the right-moving one,
\bea
K^+(L) &=& K^+_R(L) + K^+_L(L) = (1 + e^{-i 2\beta}) K^+_R(L) \nonumber\\
&=& -e^{-i \Theta} K^+(0).
\label{eq:a40}
\eea
Therefore
\bea
\label{eq:a41}
K^+(L) K^-(0) &=& -e^{-i \Theta} K^+(0) K^-(0) \nonumber \\
&=&  -4 \cos^2(\beta) \, e^{-i \Theta} K_R^+(0) K_R^-(0) \nonumber\\
&=& -\cos^2(\beta) \, e^{-i \Theta} \frac{\kappa}{2\pi\alpha}.
\eea
This shows an unusual long-ranged end-to-end correlations between spin currents at the opposite ends of the wire.

\begin{figure}
       \includegraphics[width=0.91\columnwidth]{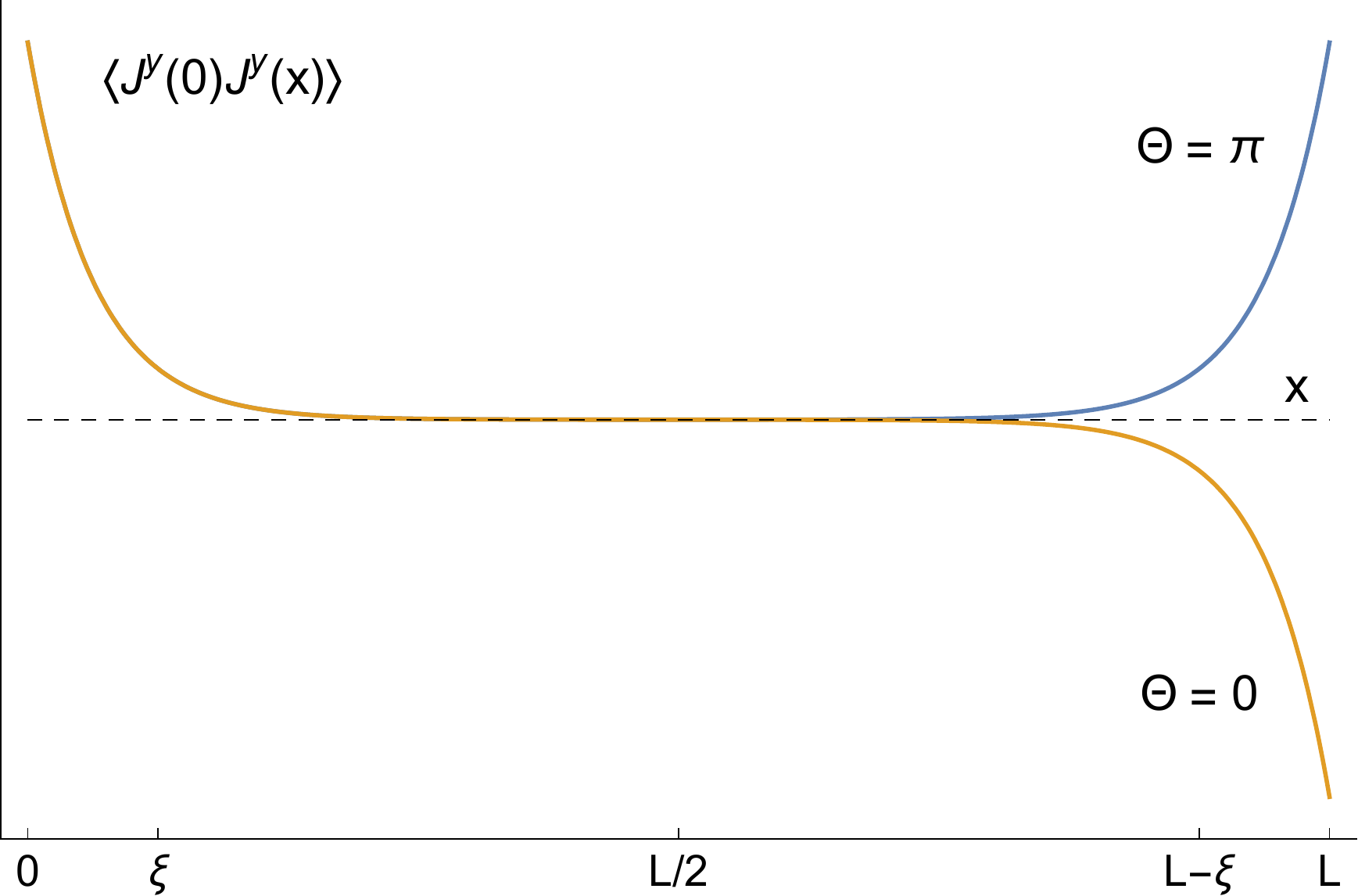}
    \caption{Schematics of the spin correlations in the SDW phase of the open quantum wire. The $\hat{y}$-component of spin density exhibits long-ranged edge-to-edge correlations, 
    Eq.~\eqref{eq:a42}, while the edge-to-bulk correlations decay exponentially on the scale $\xi=v_\sigma/\Delta$, as discussed in Sec.\ref{sec:bulk}. Staggered components of the spin density, $N^y$, behave similarly.}
     \label{fig:end-to-end}
\end{figure}

Equation \eqref{eq:a41} is to be contrasted with exponentially decaying correlations of $K^+_R$ in the bulk of the wire, as discussed in Sec.~\ref{sec:bulk} above.
The strength of the correlation between the opposite edges is determined by the spin gap, $\kappa \approx \Delta/v_\sigma$, see \eqref{eq:o20}.

Spin currents at the ends of the wire in the original basis are related to $K^a$ via \eqref{Rotation}, when position-dependent rotation \eqref{eq:ca22} reduces to matrix identity.
Therefore at $x_o$ we obtain
\bea
J^+_{R/L}(x_o) = e^{\mp i\beta} K^+_{R/L}(x_o)
\label{eq:46}
\eea
and, correspondingly, find the original spin currents at the ends of wire are proportional to Majorana mode $d_2$ as well,
\bea
J^+(0) &=& J^+_R(0) + J^+_L(0) = 2 e^{-i \beta} K^+_R(0) = \frac{2}{\sqrt{2\pi\alpha}} f^\dagger(0), \nonumber\\
J^+(L) &=& -2 e^{-i\Theta} e^{-i\beta} K^+_R(0) = \frac{-2 e^{-i\Theta}}{\sqrt{2\pi\alpha}} f^\dagger(0),
\label{eq:a53}
\eea
where
\be
f^\dagger(0)=-i\sqrt{\frac{\kappa}{2}}d_2.
\ee
Since $\Theta = 0$ or $\pi$, the above equation means that $J^x(x_o) = 0$ while the $\hat{y}$-component of $\vec{J}(x_o)$ is finite, $J^y(x_o) \sim d_2$. 
The end-to-end correlation of the uniform part of the spin density are similarly long-ranged,
\bea
\label{eq:a42}
J^+(L) J^-(0) = -e^{-i \Theta}\frac{ \kappa}{2\pi\alpha}  = - J^+(L) J^+(0). 
\eea
It differs from the same-position correlation of the spin density at the end of the wire only by the sign, $J^+(0) J^-(0) = \kappa/(2\pi\alpha)$. 
The last equality in \eqref{eq:a42} reflects the fact that $J^+(x_o) = i J^y(x_o)$, as noted above. This behavior is schematically sketched in Figure~\ref{fig:end-to-end}. 

Next we look at the correlations of the staggered part $\vec{N}$ of the spin density. Equation \eqref{x5} shows that near the ends of the wire
staggered spin density is also proportional to Majorana modes
\bea
N^+(x_o) &\propto& e^{-i \sqrt{2\pi} [\Phi_{R \rho}(x_o) - \Phi_{R \rho}(-x_o)]}  e^{-i \sqrt{\pi} [\tilde{\Phi}(x_o) + \tilde{\Phi}(-x_o)]} \nonumber\\
&=& e^{-\sqrt{4\pi} \tilde{\Phi}(x_o)} \sim f^\dagger(x_o)
\label{x8}
\eea
because both $\Phi_{R \rho}$ and $\tilde{\Phi}$ are $2L$-periodic. 

As a matter of fact, it is easy to argue that end-to-end correlations of $N^+$ field must be identical to those of $J^+$ one, \eqref{eq:a42}.
This is because at the ends of the open wire the total spin density must vanish, and therefore 
\be
\vec{S}(x_o) = \vec{J}_R(x_o) +  \vec{J}_L(x_o) 
+ [\vec{N}(x_o) e^{-i 2k_F x_o} + {\rm h.c.}] = 0,
\label{eq:a51}
\ee
since in addition $e^{\pm i 2k_F x_o} = 1$. Hence indeed, staggered components of the spin density 
possess the same end-to-end correlations as the uniform ones, \eqref{eq:a42}.
Technically, this happens because near the wire's ends the charge exponential in \eqref{x8} can be expanded as $1-i\sqrt{8\pi} \alpha d\Phi_{R\rho}/dx+...$ and 
be approximated by the unity.  That is, near the wire's ends charge fluctuations are frozen out, while the spin part of $N^+$ reduces to the negative of $J^+$ one at the same time.
Explicit calculation based on full expressions given in Appendix \ref{app:observe} confirms this natural conclusion.

Once again, we see that spin density exhibits an unusual long-range end-to-end correlations despite the fact that in the bulk of the wire all correlations decay, some exponentially fast (such as spin currents and $N^{0,z}$) while others algebraically ($N^{\pm}$), as discussed in Sec.~\ref{sec:bulk}.

 \subsection{Two-fold degeneracy and the magnetization parity}
 \label{sec:spt}
 
Proportionality of $\vec{J}$ and $\vec{N}$ to the fermion operators $f^\dagger(0)$ and $f^\dagger(L)$ merits additional discussion. Consider the wire with a
fixed total number of electrons $N_\rho = N_\uparrow + N_\downarrow$. Magnetization is $M = (N_\uparrow - N_\downarrow)/2$. Therefore parities of 
spin-$\uparrow$ and $\downarrow$ band are
$(-1)^{N_\uparrow} = e^{i \pi N_\rho/2} e^{i \pi M}$ and $(-1)^{N_\downarrow} = e^{i \pi N_\rho/2} e^{-i \pi M}$. It is sufficient to discuss just one of them, say 
$P_1 = (-1)^{N_\uparrow}$. Let us assume, for definiteness, that $N_\rho$ is {\em even}, so that the factor $e^{i \pi N_\rho/2}$ is real-valued.
Then $2M$ is also even and 
hence $e^{i \pi M}$ has eigenvalues $\pm 1$. As discussed below \eqref{eq:magparity}, Hamiltonian ${\widetilde{\cal H}}_\sigma$ conserves magnetization parity $e^{i \pi M}$, and therefore the parity $P_1$ as well. Hence the ground state of ${\widetilde{\cal H}}_\sigma$ is characterized by the definite parity $P_1$. 
But anti-commutation of $e^{i \pi M}$ and $f^\dagger(x)$, Eq.~\eqref{eq:a35}, implies that the expectation value of $f^\dagger(x_o)$ in the state $|\phi\rangle$ with definite 
fermion parity is zero. Indeed,
\be
0 = \langle \phi| e^{i \pi \hat{M}} f^\dagger(x_o) + f^\dagger(x_o) e^{i \pi \hat{M}} | \phi\rangle = 2 e^{i \pi M} \langle \phi| f^\dagger(x_o)|\phi\rangle ,
\label{x9}
\ee
where $e^{i \pi M}$ on the right hand side of \eqref{x9} is the eigenvalue of the parity operator $e^{i\pi \hat{M}}$ in the state $|\phi\rangle$, where we have used $\hat{M}$ for the magnetization operator $M$.
Therefore we conclude that the ground state expectation value of spin operators $\vec{J}$ and $\vec{N}$ near the ends of the wire is zero, 
$\langle \phi| J^+(x_o)|\phi\rangle = 0 = \langle \phi| N^+(x_o) |\phi\rangle$. And, as discussed above, expectation value of operators 
$\vec{J}$ and $\vec{N}$ in the bulk of the wire is zero, too. 

We thus see that the ground state of the Hamiltonian \eqref{eq:34} is disordered but two-fold degenerate. The degeneracy is topological, it comes from the degeneracy of 
many-body ground states $|0\rangle$ and $|1\rangle$, which have opposite magnetization parities. 
These states are defined via $\gamma_0 |0\rangle = 0$ and $|1\rangle = \gamma_0^\dagger |0\rangle$. Let $|0\rangle$ be an eigenstate of the parity $e^{i \pi \hat{M}}$ with eigenvalue 
$e^{i \pi M} = \pm 1$ , that is 
\be
e^{i \pi \hat{M}} |0\rangle = e^{i \pi M} |0\rangle .
\label{x10}
\ee
Then state $|1\rangle$ has the opposite parity,
\be
e^{i \pi \hat{M}} |1\rangle = e^{i \pi \hat{M}} \gamma_0^\dagger|0\rangle = - \gamma_0^\dagger e^{i \pi M} |0\rangle = - e^{i \pi M} |1\rangle .
\label{x11}
\ee
Note that these two states share property \eqref{x9}, that is $\langle 0| f(x_o)|0\rangle = \langle 1| f(x_o)|1\rangle = 0$.

Conversely, we can construct states $|u/d\rangle = (|0\rangle \pm |1\rangle)/\sqrt{2}$, where the plus (minus) sign corresponds to the state $|u\rangle$ ($|d\rangle$), for which the expectation value of $J^+, N^+$ operators near the ends is finite, 
\bea
\langle u/d | \gamma_0 |u/d \rangle
&=&
\frac{1}{2}(\langle 0| \pm \langle 1|)
(\gamma_0 |0\rangle \pm \gamma_0|1\rangle)
= \pm \frac{1}{2} \langle 0| \gamma_0 |1\rangle
\nonumber\\
&=&
\pm \frac{1}{2}, \\
\langle u/d | \gamma_0^\dagger | u/d \rangle
&=&
\frac{1}{2}(\langle 0| \pm \langle 1|) \gamma_0^\dagger |0\rangle = \pm \frac{1}{2}.
\nonumber
\eea
But for these states magnetization-parity is not defined
\be
e^{i\pi\hat{M}} |u/d\rangle = \frac{e^{i\pi M}}{\sqrt{2}} (|0\rangle - (\pm) |1\rangle) = e^{i\pi M} |d/u\rangle .
\ee
Rather, the parity operator $e^{i\pi\hat{M}}$ represents Pauli matrix $\sigma^x$ in the subspace spanned by the states $|u\rangle$ and $ |d\rangle$.

The physical states of the open wire are of the type $|0\rangle, |1\rangle$ from the above, simply because they are characterized by the definite magnetization parity.

Fermi operator $\gamma_0$ is introduced in \eqref{eq:a32} and, according to the discussion above, can be constructed with the help of equations 
\eqref{eq:o23}, \eqref{eq:o24}, \eqref{x7} and \eqref{eq:46} as
\be
\gamma_0 = -\sqrt{\frac{2\pi\alpha}{\kappa}}\Big( J^+_R(0) - e^{i \Theta/2} e^{i \pi \hat{M}} J^+_R(L)\Big) ,
\label{x12}
\ee
which makes explicit its non-local nature.

It is also useful to notice now that the single fermion operator \eqref{eq:a7} does not have simple expression in terms of $f$ because
\be
e^{i \sqrt{2\pi} \Phi_{R \sigma}(x)} \sim e^{i \sqrt{\pi}[\tilde{\Phi}(x) + \tilde{\Phi}(-x)]/2 + i\sqrt{\pi}[\tilde{\Phi}(x) - \tilde{\Phi}(-x)]}
\label{eq:54}
\ee
and therefore reduces to the ``square root" of Majorana in the $x\to x_o$ limit, e.g. $\psi_R(0) \sim e^{i\sqrt{\pi} \tilde{\Phi}(0)} \sim [f(0)]^{1/2}$.
At the same time, in the bulk of the wire the presence of the dual combinations $\tilde{\Phi}(x) + \tilde{\Phi}(x)$ and $\tilde{\Phi}(x) - \tilde{\Phi}(-x)$
in \eqref{eq:54} imply exponential decay of correlation functions of fermion operator $\psi_R$ with distance and time. That is, gapped behavior, just as expected.

\subsection{Instability of the two-fold degeneracy to the parity-breaking perturbations}
\label{sec:pert}

There are several physically-reasonable perturbations which violate the magnetization parity conservation \cite{Cheng2011}.

(i) Local magnetic field near the end of the wire, such as, for example, due to the magnetic impurity $\delta H' = \vec{h}_1 \cdot \vec{S}(a)$, located at a short
distance $a < \xi = v_\sigma/\Delta$ from, for example, the left end of the wire. Then $a/\xi < 1$ ensures that $\vec{h}_1$ couples the end-modes of the wire. Using 
$\vec{S}(x) = \vec{J}(x) + [\vec{N}(x) e^{-i 2k_F x} + {\rm h.c.}]$ and the fact that $2 \vec{N}(a) \approx - \vec{J}(a)$, we find that 
$\delta H_1 \approx [1-\cos(2k_F a)] \vec{h}_1 \cdot \vec{J}(a)$.
Now, the discussion around \eqref{eq:a53} and \eqref{eq:a42} shows that 
$\vec{J}(a) \approx i \sqrt{\kappa/(2\pi\alpha)} e^{-\kappa a} (\gamma_0 - \gamma_0^\dagger) \hat{y}$, so that we can write
$\delta H_1 = i \tilde{h}_1 (\gamma_0 - \gamma_0^\dagger)$ by absorbing all non-essential constants into $\tilde{h}_1$. In the low-energy subspace of definite parity states 
$\{|0\rangle, |1\rangle \}$  this perturbation is off-diagonal and reduces to $\delta H' = \tilde{h}_1 \sigma^y$. Its eigenstates are those of the Pauli matrix 
$\sigma^y$, with energies $\pm \tilde{h}_1$, and they are {\em not} eigenstates of the magnetization parity $e^{i\pi\hat{M}}$.

Therefore this local perturbation breaks magnetization parity conservation and removes the two-fold degeneracy of the ground state in favor 
of the unique state $(|0\rangle - i |1\rangle)/\sqrt{2}$ with energy $-\tilde{h}_1$.

An interesting consequence of the end-to-end correlation \eqref{eq:a42}, which for $2M={\rm even}$ (see beginning of this Section where $N_\rho={\rm even}$ was set) 
can be written as $J^y(L) J^y(a) \approx -\kappa/(2\pi\alpha)$, is that $\delta H'$ acting near the left end of the wire causes finite polarization $J^y(L)$ at its opposite, right end.
This kind of ``long-distance rigidity'' in the absence of rigid spin correlations in the bulk of the wire is unusual and represents a bosonic version of the teleportation phenomena previously suggested for fermion Majorana states \cite{tewari2008,Fu2010}.

(ii) Next, consider applying the local magnetic field somewhere in the middle of the wire, so that the perturbation still has the form $\delta H'$ but now with $\xi \ll a \ll L-\xi$.
The uniform part of the spin density is exponentially suppressed there and the field couples to the staggered part, 
$\widetilde{N}(a)$. Since $\tilde{\Phi}(x) + \tilde{\Phi}(-x)$ is locked to the optimal value, 
$\delta H'$ reduces to 
\bea
\delta H' &\approx&
(h_{1x}-i h_{1y}) {\cal F}^\dagger e^{- i\frac{\pi x}{L}(N_\rho +1)}
e^{-i\sqrt{2\pi}[\Phi_{R \rho}(x) - \Phi_{R \rho}(-x)]}
\nonumber\\ &&
 + {\rm h.c.},
\label{eq:a53b}
\eea
see \eqref{x5}. The expectation value of this operator in a finite-size system is proportional to $(\alpha/L)^{K_\rho/2}$. 
This is because projection of the charge-mode exponential to the ground state gives, after normal-ordering it,
\be
e^{-i\sqrt{2\pi}[\Phi_{R \rho}(x) - \Phi_{R \rho}(-x)]} \to
 \left(\frac{\pi \alpha}{2 L |\sin(\pi x/L)|}\right)^{K_\rho/2}
\ee
for $\alpha \ll x \ll L$.
Note also that fermion changing operator ${\cal F}$ in \eqref{eq:a53} violates the conservation of the magnetization parity.
Therefore such a perturbation, which has the meaning of electron spin-flip backscattering off a magnetic impurity, also breaks the two-fold degeneracy of the ground state.
Here the breaking of the degeneracy due to the perturbation is smaller than in the previous example (i), it vanishes algebraically with the size of the 
system as $h_1 (\alpha/L)^{K_\rho/2}$, and therefore is not particularly important for sufficiently long wires.

\section{Discussion}
\label{sec:disc}

We found that ground states of the interacting wire in the correlated SDW phase has all the features of the symmetry-protected topological (SPT) state \cite{turner2011,xiechen2011,schuch2011}. Indeed, in the closed (ring) geometry the ground state is unique and is an eigenstate of a definite magnetization parity $e^{i \pi N_\sigma/2}$. In the open wire geometry, however, the ground states corresponding to different magnetization parities $e^{i \pi M}$ are degenerate with exponential accuracy $e^{-L/\xi}$. The localized Majorana modes that appear in this geometry are found to describe spin density near the wire's ends. Importantly, the expectation value of the spin density 
in the state with definite magnetization parity is zero everywhere in the wire, including its ends, and can not be used to distinguish the degenerate ground states. 
The correlated SDW state can therefore be classified as a SPT state which is protected by the magnetization parity.

In the topological SDW state the spin sector is Ising ordered along the $y$ direction (the direction of the Rashba spin-orbit interaction) while the charge sector is a 
gapless Tomonaga-Luttinger liquid. The charge fluctuations weaken the correlation of the $S^y$ spins and make it quasi-long-ranged in the bulk.
However, at the ends of a wire the charge fluctuations are frozen so that the long-range Ising spin correlation between the end spins can manifest itself without 
being obscured by charge fluctuations. In some sense this long-range spin correlation is just the Ising order of the XYZ spin chain covered by the critical charge fluctuations in the bulk.
Nevertheless, we regard the SDW state as a SPT state, because its ground state degeneracy is determined by the boundary conditions -- the ground state is unique in the ring geometry with both PBC and anti-PBC conditions and becomes two-fold degenerate in the open wire geometry. The degeneracy is protected by the magnetization parity. Indeed, it takes a parity-breaking perturbation to lift the ground state degeneracy, as we show in Section \ref{sec:spt}.

It is important to note that without critical charge mode our model would reduce to the transverse field Ising (TFI) chain the ground state of which is not an SPT phase. This is seen from the fact that in the ring geometry the TFI model retains two-fold degeneracy (with exponentially small in the system size splitting) which is just the usual Ising $Z_2$ degeneracy. In our case it is the gapless charge mode which endows gapped quantum wire with the SPT properties.

This observation is in agreement with several previous studies of topological states of interacting quantum wires \cite{Cheng2011,keselman2015,kainaris2015,kainaris2017}
which found that the presence of the gapless charge (more generally, center-of-mass) mode is crucial for the ground state degeneracy.
It must be added here that as far as possible physical realization of the described physics goes, the model studied here appears to be the simplest one.
Its realization requires only a quantum wire with significant spin-orbit coupling and strong repulsive e-e interactions -- and all of these ingredients are readily available in the present-day experimental setups.

Another important lesson of our study follows from the fact that it is collective spin degrees of freedom, which are described by boson-like operators, 
that become `topologically' correlated. 
The difference between a one-particle fermion operator and a two-particle boson operator is fundamental.
Simple one-particle fermion operators at different points must anti-commute, and naturally they do, as \eqref{eq:a32} and \eqref{eq:a33} show. 
The two-particle operators, which necessarily are boson-like, such as the spin density here,
on the other hand, must commute when taken at different points. This is achieved with the help of the string operator $e^{-i\pi M}$, as explained in Section \ref{sec:end-end}. 
Therefore two-particle operators at the opposite ends of the wire {\em must} be proportional to each other, up to unessential phase factors. 
This is the crucial difference between the fundamental degrees of freedom of the fermionic Kitaev chain (one-dimensional $p$-wave superconductor), which are single-particles of BdG type \cite{Kitaev2001,loss2020},  and the transverse field Ising chain, where they are two-particle excitations of magnetic kind, and we have re-discovered it here for the correlated SDW wire.

We therefore arrive at the logical conclusion that many-body states are not particularly good for realizing Majorana degrees of freedom as long as they are based on some kind of 
two-particle (bose-like) operators. For single-particle based constructions, such as weakly-interacting semiconducting quantum wires in contact with a superconductor, 
the commutation requirement does not exist and therefore there are no fundamental restrictions to realizing the sought-after Majorana fermions in such platforms.  

We nonetheless believe that our problem is interesting in its own way. It shows how fractionalized degrees of freedom emerge in a basic set-up of a quantum wire with repulsive interactions only. The finding of the two-fold degenerate ground state with unusual long-ranged correlations between the spin densities at the opposite ends of the open wire, while the correlations in the bulk of the wire decay, at best, as a power-law, deserves further theoretical and experimental studies.

\begin{acknowledgments}

We would like to thank numerous people discussions with whom have contributed significantly to our understanding of the subject: Jason Alicea, Patrick Azaria, Leon Balents,  Erez Berg, Piet Brouwer, Sam Carr, Jan von Delft, Thierry Giamarchi, Leonid Glazman, Henrik Johannesson, Anna Keselman, Daniel Loss, Christopher Mudry, Yuval Oreg.
We thank \"Omer Aksoy and Christopher Mudry for helpful comments on the manuscript.

The work of AF was supported by JSPS KAKENHI (Grant No.\ 19K03680) and JST CREST (Grant No.\ JPMJCR19T2). The work of OAS and RBW was supported by the NSF CMMT program under grant DMR-1928919. OAS thanks FY2017 JSPS Invitational Fellowship for Research in Japan (award No. S17117) which supported his visit to RIKEN where this work was initiated.
\end{acknowledgments}

\appendix

\section{Screened Coulomb interaction and spin-charge separation}
\label{sec: interaction and spin-charge separation}

In this Appendix we derive the low-energy effective theory
in terms of current operators $J_{R/L}(x)$ and $\vec{J}_{R/L}(x)$.
The key idea of the derivation is similar to \cite{capponi2000,maslov2005} although on a technical level we proceed by employing operator-product-expansion (OPE) of fermion operators while these references applied them in the bosonic language.
We assume the limit $L\to\infty$ in this Appendix.
We use simplified notations
\be
R_s(x)=\Psi_{Rs}(x),\quad
L_s(x)=\Psi_{Ls}(x),
\ee
and write the electron density as
\begin{subequations}
\be
\rho(x)=
\sum_{s=\uparrow,\downarrow}
:\!\Psi^\dagger_s(x)\Psi_s^{}(x)\!:
\, = \rho_0(x)+\rho_{2k_F}(x),
\ee
where
\bea
\rho_0(x)
&=&
\sum_{s=\uparrow,\downarrow}\!\left[
:\!R_s^\dagger(x)R_s^{}(x)\!: \!
+ \! :\!L_s^\dagger(x)L_s^{}(x)\!:
\right]
\nonumber\\
&=&
J_R(x)+J_L(x),
\\
\rho_{2k_F}(x)
&=&
\sum_{s=\uparrow,\downarrow}\!\left[
\,e^{2ik_Fx}L_s^\dagger(x)R_s^{}(x)
+e^{-2ik_Fx}R_s^\dagger(x)L_s^{}(x)
\right]\!\! .
\nonumber\\&&
\eea
\end{subequations}
The density-density interaction is decomposed into two parts
\begin{equation}
H_\mathrm{int}=
\frac12\int\!dx\!\int\!dy\,\rho(x)U(x-y)\rho(y)
\approx
H_F+H_B,
\end{equation}
where forward- and backward-scattering interactions are
\begin{eqnarray}
H_F&=&
\frac12
\int\!dx\!\int\!dy \, \rho_0(x)U(x-y)\rho_0(y),
\label{H_F spinful}
\\
H_B&=&
\frac12
\int\!dx\!\!\int\!dy \, \rho_{2k_F}(x)U(x-y)\rho_{2k_F}(y).
\label{H_B spinful}
\end{eqnarray}
Here we have discarded rapidly oscillating cross terms ($\rho_0\rho_{2k_F}$).

The interaction potential $U(x-y)$ is short-ranged.
Introducing new variables $X=(x+y)/2$ and $\tilde{x}=x-y$,
we can rewrite the forward-scattering interaction $H_F$ as
\begin{eqnarray}
H_F&\approx&
\frac12\int\!d\tilde{x}\,U(\tilde{x})\int\!dX
\left[\rho_0(X)\right]^2
\nonumber\\
&=&
\frac{U_0}{2}\int dx\,[J_R(x)+J_L(x)]^2.
\end{eqnarray}

The product of $\rho_{2k_F}$ in $H_B$ yields
\bea
\rho_{2k_F}(x)\rho_{2k_F}(y)
&\approx&
e^{2ik_F\tilde{x}}\sum_{\mu,\sigma}
L_\mu^\dagger(x)R_\mu^{}(x)R_\sigma^\dagger(y)L_\sigma^{}(y)
\nonumber\\
&&{}\!\!\!\!
+
e^{-2ik_F\tilde{x}}\sum_{\mu,\sigma}
R_\mu^\dagger(x)L_\mu^{}(x)L_\sigma^\dagger(y)R_\sigma^{}(y),
\nonumber\\&&
\eea
where we have discarded rapidly oscillating terms ($\propto e^{\pm 4ik_FX}$).
The backward-scattering interaction $H_B$ can be calculated
using the operator-product expansions
\begin{eqnarray}
&&
L_\mu^\dagger(x)R_\mu^{}(x)R_\sigma^\dagger(y)L_\sigma(y)
\nonumber\\
&&=\!
\left[-\frac{i\delta_{\mu,\sigma}}{2\pi\tilde{x}}
 +:\!L_\mu^\dagger(x)L_\sigma^{}(y)\!:\right]
\!
\left[\frac{i\delta_{\mu,\sigma}}{2\pi\tilde{x}}
 -:\!R_\sigma^\dagger(y)R_\mu^{}(x)\!:\right]
\nonumber\\
&&=
\frac{\delta_{\mu,\sigma}}{(2\pi\tilde{x})^2}
-:\!L_\mu^\dagger(X)L_\sigma^{}(X)\!: :\!R_\sigma^\dagger(X)R_\mu^{}(X)\!:
\nonumber\\
&&{}~
+\frac{i\delta_{\mu,\sigma}}{2\pi\tilde{x}}\!\left[\,
:\!L_\sigma^\dagger(X)L_\sigma^{}(X)\!:
+:\!R_\sigma^\dagger(X)R_\sigma^{}(X)\!:\,\right]
\nonumber\\
&&{}~
+\frac{i\delta_{\mu,\sigma}}{4\pi}\left\{
\,:\![\partial_X L_\sigma^\dagger(X)]L_\sigma^{}(X)\!:
 -:\!L_\sigma^\dagger(X)\partial_XL_\sigma^{}(X)\!:
\right.\nonumber\\&&\left.\qquad\qquad
-:\![\partial_X R_\sigma^\dagger(X)]R_\sigma^{}(X)\!:
 +:\!R_\sigma^\dagger(X)\partial_XR_\sigma(X)\!:\,
\right\}
\nonumber\\&&{}~
+\ldots,
\label{OPE1 spinful}
\end{eqnarray}
and
\begin{eqnarray}
&&
R_\mu^\dagger(x)L_\mu^{}(x)L_\sigma^\dagger(y)R_\sigma^{}(y)
\nonumber\\
&&=\!
\left[-\frac{i\delta_{\mu,\sigma}}{2\pi\tilde{x}}
-:\!L_\sigma^\dagger(y)L_\mu^{}(x)\!:\right]\!
\left[\frac{i\delta_{\mu,\sigma}}{2\pi\tilde{x}}
+:\!R_\mu^\dagger(x)R_\sigma^{}(y)\!:\right]
\nonumber\\
&&=
\frac{\delta_{\mu,\sigma}}{(2\pi\tilde{x})^2}\,
-:\!L_\sigma^\dagger(X)L_\mu^{}(X)\!: :\!R_\mu^\dagger(X)R_\sigma^{}(X)\!:
\nonumber\\
&&{}~
-\frac{i\delta_{\mu,\sigma}}{2\pi\tilde{x}}\left[\,
:\!L_\sigma^\dagger(X)L_\sigma^{}(X)\!:
+:\!R_\sigma^\dagger(X)R_\sigma^{}(X)\!:\,\right]
\nonumber\\
&&{}~
+\frac{i\delta_{\mu,\sigma}}{4\pi}\!\left\{
\,:\![\partial_X L_\sigma^\dagger(X)]L_\sigma^{}(X)\!:
-:\!L_\sigma^\dagger(X)\partial_XL_\sigma^{}(X)\!:
\right.\nonumber\\&&\left.\qquad\qquad
-:\![\partial_X R_\sigma^\dagger(X)]R_\sigma^{}(X)\!:
+:\!R_\sigma^\dagger(X)\partial_XR_\sigma^{}(X)\!:\,
\right\}
\nonumber\\&&{}~
+\ldots.
\label{OPE2 spinful}
\end{eqnarray}
The second term in the above expansions can be written
in terms of currents,
\bea
\sum_{\mu,\sigma=\uparrow,\downarrow}\!\!
:\!L_\mu^\dagger(x)L_\sigma^{}(x)\!:\,:\!R_\sigma^\dagger(x)R_\mu^{}(x)\!:
\,
&=&
\frac{1}{2}J_L(x)J_R(x)
\nonumber\\
&&{}
+2\vec{J}_L(x)\cdot\vec{J}_R(x),
\nonumber\\&&
\eea
and the fourth term (a kinetic energy density)
can be written as
\bea
&&
-\frac{i}{2}\!\left\{
\,:\![\partial_x L_\sigma^\dagger(x)]L_\sigma^{}(x)\!:
-:\!L_\sigma^\dagger(x)\partial_xL_\sigma^{}(x)\!:
\right.\nonumber\\&&\left.\qquad
-:\![\partial_x R_\sigma^\dagger(x)]R_\sigma^{}(x)\!:
+:\!R_\sigma^\dagger(x)\partial_xR_\sigma^{}(x)\!:\,
\right\}
\nonumber\\
&&=
\frac{\pi}{2}\!\left\{:\![J_L(x)]^2\!:+:\![J_R(x)]^2\!:\right\}
\nonumber\\&&{}\quad
+\frac{2\pi}{3}\!\left[:\!\vec{J}_L(x)\cdot\vec{J}_L(x)\!:
+:\!\vec{J}_R(x)\cdot\vec{J}_R(x)\!:\right] .
\eea
Combining these contributions, we obtain
\bea
H_B&=&
\int\!dx\int\!d\tilde{x}\,U(\tilde{x})\frac{\cos(2k_F\tilde{x})}{2\pi^2\tilde{x}^2}
\nonumber\\
&&\!
-\int\!d\tilde{x}\,U(\tilde{x})\frac{\sin(2k_F\tilde{x})}{2\pi\tilde{x}}
\!\int\!\!dx\left[J_L(x)+J_R(x)\right]
\nonumber\\
&&{}\!\!
-U_{2k_F}\!
\int\!dx\,\biggl\{
\frac{1}{2}J_L(x)J_R(x)+2\vec{J}_L(x)\cdot\vec{J}_R(x)
\nonumber\\
&&{}\hspace*{16mm}
+\frac{1}{4}\!\left\{:\![J_L(x)]^2+[J_R(x)]^2\!:\right\}
\nonumber\\
&&{}\hspace*{16mm}
+\frac{1}{3}\!\left[:\!\vec{J}_L(x)\!\cdot\!\vec{J}_L(x)\!:
+:\!\vec{J}_R(x)\!\cdot\!\vec{J}_R(x)\!:\right]\!
\biggr\},
\nonumber\\&&
\label{H_B after OPEs spinful}
\eea
where
\be
U_{2k_F}=\int\!d\tilde{x}\,U(\tilde{x})\cos(2k_F\tilde{x}).
\ee
The first term on the right-hand side of \eqref{H_B after OPEs spinful}
is a constant, and the second term is renormalization of the chemical potential.
We thus keep the last contributions proportional to $U_{2k_F}$ and
finally obtain
\be
H_F+H_B=H_c+H_s+\ldots,
\ee
with the charge part
\be
H_c=\frac{1}{4}(2U_0-U_{2k_F})\int\!dx\,[J_R(x)+J_L(x)]^2,
\label{x13}
\ee
and the spin part
\bea
H_s&=&
-2U_{2k_F}\int\!dx\,\vec{J}_L(x)\cdot\vec{J}_R(x)
\nonumber\\
&&{}
-\frac{U_{2k_F}}{3}\int\!dx\,
\left[:\!\vec{J}_L(x)\!\cdot\!\vec{J}_L(x)\!:
+:\!\vec{J}_R(x)\!\cdot\!\vec{J}_R(x)\!:\right].
\nonumber\\&&
\label{H_s}
\eea
The second line in \eqref{H_s} gives renormalization of
the velocity $v_F$ in \eqref{eq:6}.

We note that $H_c$ is a functinal of $J_R+J_L$, which is a consequence
of Galilean invariance.

\section{Bosonization}

\subsection{Bosonization under periodic boundary condition}
\label{sec: bosonization periodic}

Here we summarize bosonization rules for the ring geometry \cite{Eggert1992,Wong1994,Oshikawa2006}.
We first define chiral bosonic fields
\begin{subequations}
\label{chiral phi}
\bea
\phi_{Rs}(x)&=&\phi_{Rs}^0+\frac{\sqrt{\pi} x}{L}N_{Rs}+\varphi_{Rs}(x),\\
\phi_{Ls}(x)&=&\phi_{Ls}^0+\frac{\sqrt{\pi} x}{L}N_{Ls}+\varphi_{Ls}(x),
\eea
\end{subequations}
where $s=\uparrow,\downarrow$, and
\be
[\phi_{Rs}(x),\phi_{Rs'}(y)]
=-[\phi_{Ls}(x),\phi_{Ls'}(y)]
=\frac{i}{4}\delta_{s,s'}\mathrm{sgn}(x-y).
\ee
The zeromode operators satisfy the commutation relations
\bea
&&
[\phi_{Rs}^0,N_{Rs'}]=-[\phi_{Ls}^0,N_{Ls'}]=-\frac{i\delta_{s,s'}}{\sqrt{4\pi}},
\\
&&
[\phi_{Rs}^0,N_{Ls'}]=[\phi_{Ls}^0,N_{Rs'}]=[\phi_{Rs},\phi_{Ls'}]=[N_{Rs}.N_{Ls'}]=0.
\nonumber\\&&
\eea
The fields $\varphi_{Rs}$ and $\varphi_{Ls}$ are periodic functions of $x$,
$\varphi_{R/Ls}(x+L)=\varphi_{R/Ls}(x)$, and can be expanded as
\bea
\varphi_{Rs}(x)&=&\sum^\infty_{n=1}
\frac{e^{-\pi\alpha n/L}}{\sqrt{4\pi n}}(e^{iq_nx}a_{n,Rs}+e^{-iq_nx}a_{n,Rs}^\dagger),
\quad
\\
\varphi_{Ls}(x)&=&\sum^\infty_{n=1}
\frac{e^{-\pi\alpha n/L}}{\sqrt{4\pi n}}(e^{-iq_nx}a_{n,Ls}+e^{iq_nx}a_{n,Ls}^\dagger),
\eea
where $q_n=2\pi/L$,
$\alpha$ is a shorot-distance cutoff, and the boson annihilation/creation operators obey the commutation relations
\be
[a_{n,Rs}^{},a_{n',Rs'}^\dagger]=[a_{n,Ls}^{},a_{Ls'}^\dagger]=\delta_{n,n'}\delta_{s,s'}.
\ee
The fields with different chiralities commute,
\be
[\phi_{Rs}(x),\phi_{Ls'}(y)]=0.
\ee

The chiral fermion fields are written in terms of the chiral bosonic fields as follows:
\bea
\psi_{Rs}(x)&=&\frac{\eta_{Rs}}{\sqrt{2\pi\alpha}}
e^{i\sqrt{4\pi}\phi_{Rs}^0}e^{2\pi iN_{Rs}x/L+i\sqrt{4\pi}\varphi_{Rs}(x)}
\nonumber\\
&=&
\frac{\eta_{Rs}}{\sqrt{2\pi\alpha}}
e^{i\sqrt{4\pi}\phi_{Rs}(x)+i\pi x/L},
\\
\psi_{Ls}(x)&=&\frac{\eta_{Ls}}{\sqrt{2\pi\alpha}}
e^{-i\sqrt{4\pi}\phi_{Ls}^0}e^{-2\pi iN_{Ls}x/L-i\sqrt{4\pi}\varphi_{Ls}(x)}
\nonumber\\
&=&
\frac{\eta_{Ls}}{\sqrt{2\pi\alpha}}
e^{-i\sqrt{4\pi}\phi_{Ls}(x)-i\pi x/L},
\eea
where $\eta_{R/Ls}$ obey the anticommutation relations
\be
\{\eta_{Rs},\eta_{Rs'}\}=\{\eta_{Ls},\eta_{Ls'}\}=2\delta_{s,s'},
\quad
\{\eta_{Rs},\eta_{Ls'}\}=0.
\ee
The fermion field operators $\psi_{Rs}$ and $\psi_{Ls}$ satisfy the standard
anticommutation relations.
The fermion density operators are given by
\bea
:\!\psi_{Rs}^\dagger(x)\psi_{Rs}^{}(x)\!:\,&=&\frac{1}{\sqrt\pi}\partial_x\phi_{Rs}(x),
\\
:\!\psi_{Ls}^\dagger(x)\psi_{Ls}^{}(x)\!:\,&=&\frac{1}{\sqrt\pi}\partial_x\phi_{Ls}(x),
\eea
and therefore the fermion number operators $N_{Rs}$ and $N_{Ls}$, defined by
\bea
N_{Rs}&=&\int^L_0dx\frac{1}{\sqrt\pi}\partial_x\phi_{Rs}(x),
\\
N_{Ls}&=&\int^L_0dx\frac{1}{\sqrt\pi}\partial_x\phi_{Ls}(x),
\eea
are integer-vaued operators.
The charge current operators defined in \eqref{eq:charge-current1} are
thus given by
\bea
J_R(x)&=&\frac{1}{\sqrt\pi}\partial_x[\phi_{R\uparrow}(x)+\phi_{R\downarrow}(x)],
\\
J_L(x)&=&\frac{1}{\sqrt\pi}\partial_x[\phi_{L\uparrow}(x)+\phi_{L\downarrow}(x)].
\eea
One can show, using $e^{2\pi iN_{Rs}}=e^{2\pi N_{Ls}}=1$, that the fermion fields satisfy the periodic boundary conditions,
$\psi_{Rs}(x+L)=\psi_{Rs}(x)$ and $\psi_{Ls}(x+L)=\psi_{Ls}(x)$.

The linearized kinetic energy is given by \cite{delft1998}
\bea
\int^L_0\!\!dx :\!\psi_{Rs}^\dagger(-i\partial_x)\psi_{Rs}^{}\!: 
&=&\frac{\pi}{L}(N_{Rs}^2+N_{Rs})
\nonumber\\&&\!
+\!\int^L_0\!dx :\!(\partial_x\varphi_{Rs})^2\!: ,
\\
\int^L_0\!\!dx :\!\psi_{Ls}^\dagger(i\partial_x)\psi_{Ls}^{}\!: 
&=&\frac{\pi}{L}(N_{Ls}^2+N_{Ls})
\nonumber\\&&\!
+\!\int^L_0\!dx :\!(\partial_x\varphi_{Ls})^2\!: \!.
\eea

We define nonchiral bosonic fields
\bea
\phi_s(x)=\phi_{Ls}(x)+\phi_{Rs}(x),
\\
\theta_s(x)=\phi_{Ls}(x)-\phi_{Rs}(x),
\eea
and then introduce a pair of charge field operators,
\bea
\phi_\rho(x)&=&\frac{1}{\sqrt2}
[\phi_\uparrow(x)+\phi_\downarrow(x)],
\\
\theta_\rho(x)&=&\frac{1}{\sqrt2}
[\theta_\uparrow(x)+\theta_\downarrow(x)],
\eea
and a pair of spin field operators,
\bea
\phi_\sigma(x)&=&\frac{1}{\sqrt2}
[\phi_\uparrow(x)-\phi_\downarrow(x)],
\\
\theta_\sigma(x)&=&\frac{1}{\sqrt2}
[\theta_\uparrow(x)-\theta_\downarrow(x)].
\eea
Finally, we introduce charge/spin number and current opetators,
\bea
N_\rho&=&
N_{R\uparrow}+N_{R\downarrow}+N_{L\uparrow}+N_{L\downarrow},
\label{N_rho}
\\
J_\rho&=&
N_{R\uparrow}+N_{R\downarrow}-N_{L\uparrow}-N_{L\downarrow},
\label{J_rho}
\\
N_\sigma&=&
N_{R\uparrow}-N_{R\downarrow}+N_{L\uparrow}-N_{L\downarrow},
\label{N_sigma}
\\
J_\sigma&=&
N_{R\uparrow}-N_{R\downarrow}-N_{L\uparrow}+N_{L\downarrow}.
\label{J_sigma}
\eea
By definition these operators must satisfy the following relations \cite{seidel2005}:
\bea
&&
(-1)^{N_\rho}=(-1)^{J_\rho}=(-1)^{N_\sigma}=(-1)^{J_\sigma},
\label{identity 1}
\\
&&
(-1)^{\frac{1}{2}(N_\rho+J_\rho)}=(-1)^{\frac{1}{2}(N_\sigma+J_\sigma)}.
\label{identity 2}
\eea
It is easy to write down explicit form of the spin boson $\theta_\sigma$ which will be useful for discussions in Section~\ref{sec:ring},
\bea
\theta_\sigma(x) &=& - \sqrt{\frac{\pi}{2}} \frac{x}{L} J_\sigma + \frac{1}{\sqrt{2}}\Big(\phi^0_{L\uparrow} - \phi^0_{L\downarrow} - \phi^0_{R\uparrow} + \phi^0_{R\downarrow}\nonumber\\
&& + \varphi_{L\uparrow} - \varphi_{L\downarrow} - \varphi_{R\uparrow} + \varphi_{R\downarrow} \Big).
\label{x19}
\eea
Observe that in the presence of a finite spin current $J_\sigma\neq 0$ the spin field acquires a kink at $x=0=L$ since then $\theta_\sigma(0) - \theta_\sigma(L) = \sqrt{\pi/2} J_\sigma$.

\subsection{Bosonization under open boundary condition}
\label{app:Bos-OBC}

Here we summarize bosonization rules for electrons in a wire of length $L$
with open boundaries \cite{Eggert1992,Wong1994,Hikihara1998,Hikihara2004,Oshikawa2006}.

We first define chiral boson fields
\bea
\phi_{Ls}(x)&=&
\frac{\sqrt\pi}{4}+\frac{\sqrt{\pi}x}{2L}N_s+\frac{\theta_s^0}{\sqrt{4\pi}}
+\Phi_{Ls}(x),
\label{phi_Ls open}
\\
\phi_{Rs}(x)&=&
\frac{\sqrt\pi}{4}+\frac{\sqrt{\pi}x}{2L}N_s-\frac{\theta_s^0}{\sqrt{4\pi}}
+\Phi_{Rs}(x),
\label{phi_Rs open}
\eea
where $s=\uparrow,\downarrow$,
\be
[\theta_s^0,N_{s'}]=i\delta_{s,s'},
\ee
and $\Phi_{Ls}$ and $\Phi_{Rs}$ have mode expansions,
\bea
\Phi_{Ls}(x)&=&-\Phi_{Rs}(-x)
\nonumber\\
&=&\sum^\infty_{n=1}\frac{e^{-\pi n\alpha/2L}}{\sqrt{4\pi n}}
\left(e^{-i\pi nx/L}a_{n,s}^{}+e^{i\pi nx/L}a_{n,s}^\dagger\right),
\nonumber\\&&
\label{Phi_Ls open}
\eea
which satisfy $\Phi_{Ls}(x+2L)=\Phi_{Ls}(x)$ and the same for $\Phi_{Rs}(x)$.
In \eqref{Phi_Ls open} $\alpha$ is a short-distance cutoff.

One can verify that the chiral boson fields introduced above
satisfy the commutation relations
\bea
[\phi_{Rs}(x),\phi_{Rs'}(y)]&=&
-[\phi_{Ls}(x),\phi_{Ls'}(y)]
\nonumber\\
&=&
\frac{i}{4}\delta_{s,s'}\mathrm{sgn}(x-y),
\eea
and
\be
[\phi_{Rs}(x),\phi_{Ls'}(y)]
=
\left\{\begin{array}{ll}
0, & x=y=0,\cr
{\displaystyle -\frac{i}{4}\delta_{s,s'}}, & 0<x,y<L\cr
{\displaystyle -\frac{i}{2}\delta_{s,s'}}, & x=y=L.
\end{array}\right.
\ee

We define a pair of bosonic fields ($s=\uparrow,\downarrow$)
\begin{subequations}
\bea
\phi_s(x)&=&\phi_{Ls}(x)+\phi_{Rs}(x),
\\
\theta_s(x)&=&\phi_{Ls}(x)-\phi_{Rs}(x),
\eea
\end{subequations}
which satisfy the commutation relation,
\be
[\phi_s(x),\theta_{s'}(y)]=-i\delta_{s,s'}\Theta(x-y)
\ee
for $0<x,y<L$.
The field $\phi_s(x)$ obey the Dirichelet boundary conditions at $x=0,L$:
\be
\phi_s(0)=\frac{\sqrt\pi}{2},
\quad
\phi_s(L)=\sqrt{\pi}\!\left(N_s+\frac{1}{2}\right).
\ee
We then introduce charge fields,
\begin{subequations}
\bea
\phi_\rho(x)&=&\frac{1}{\sqrt2}[\phi_\uparrow(x)+\phi_\downarrow(x)],
\\
\theta_\rho(x)&=&\frac{1}{\sqrt2}[\theta_\uparrow(x)+\theta_\downarrow(x)],
\eea
\end{subequations}
and spin fields,
\begin{subequations}
\bea
\phi_\sigma(x)&=&\frac{1}{\sqrt2}[\phi_\uparrow(x)-\phi_\downarrow(x)],
\\
\theta_\sigma(x)&=&\frac{1}{\sqrt2}[\theta_\uparrow(x)-\theta_\downarrow(x)].
\eea
\end{subequations}

Fermion fields are written in terms of the chiral boson fields
$\psi_s(x)=e^{ik_Fx}\psi_{Rs}(x)+e^{-ik_Fx}\psi_{Ls}(x)$, where \cite{fabrizio1995}
\bea
\psi_{Rs}(x)&=&
\frac{\eta_s}{\sqrt{2\pi\alpha}}e^{i\sqrt{4\pi}\phi_{Rs}(x)+i\pi x/2L}
\nonumber\\
&=&
\frac{i\eta_s}{\sqrt{2\pi\alpha}}e^{-i\theta_s^0}
e^{i\pi N_s x/L}e^{i\sqrt{4\pi}\Phi_{Rs}(x)},
\label{psi_Rs open}
\\
\psi_{Ls}(x)&=&
\frac{\eta_s}{\sqrt{2\pi\alpha}}e^{-i\sqrt{4\pi}\phi_{Ls}(x)-i\pi x/2L}
\nonumber\\
&=&
\frac{-i\eta_s}{\sqrt{2\pi\alpha}}e^{-i\theta_s^0}
e^{-i\pi N_s x/L}e^{-i\sqrt{4\pi}\Phi_{Ls}(x)}
\nonumber\\
&=&
-\psi_{Rs}(-x).
\label{psi_L = -psi_R}
\eea
Here $\eta_s$ obey the anticommutation relations
\be
\{\eta_s,\eta_{s'}\}=2\delta_{s,s'}.
\ee
The electron density operator is given by
\begin{subequations}
\bea
:\!\psi_{Rs}^\dagger(x)\psi_{Rs}^{}(x)\!:&=&\frac{1}{\sqrt\pi}\partial_x\phi_{Rs}(x),
\\
:\!\psi_{Ls}^\dagger(x)\psi_{Ls}^{}(x)\!:&=&\frac{1}{\sqrt\pi}\partial_x\phi_{Ls}(x),
\eea
\end{subequations}
We define Klein factors \cite{delft1998}
\be
F_s=\eta_s e^{-i\theta_s^0},
\ee
which satisfy
\be
F_s^\dagger F_s^{}=1,
\quad
[F_s, N_s]=F_s.
\ee 
The operator $N_s$ is integer-valued and measures the number of
electrons with spin $s$,
\be
N_s=\int^L_0\!dx
\left[
:\!\psi_{Rs}^\dagger(x)\psi_{Rs}^{}(x)\!:
+:\!\psi_{Ls}^\dagger(x)\psi_{Ls}^{}(x)\!:
\right].
\ee
It follows that $\psi_{Rs}(x+2L)=\psi_{Rs}(x)$.
The Fermi wave number is given by $k_F=\pi N_s^0/L$, where $N_s^0$ is
another integer.
We see from \eqref{psi_L = -psi_R} that
the open boundary conditions are satisfied
\be
\psi_s(0)=\psi_s(L)=0.
\ee

\section{Details of the analysis for the wire with OBC}
\label{App:openBC}

\subsection{Rotations}
\label{app:Rot}

To treat the wire with open boundaries it is convenient to orient external magnetic field along the $\hat{x}$-axis, while the spin-orbit axis continues to point along the $\hat{y}$-axis. Such a choice leads to chiral rotations about the $\hat{z}$-axis, see below, and results in convenient boundary conditions for rotated fermions $\psi(x)$, as we demonstrate now. Thus,
the Zeeman magnetic field $b = g \mu_B^{} B$ couples to the magnetization along the $\hat{x}$-axis,
\be
{\cal V}_{\rm x}  = - b \int dx~ (J^x_R + J^x_L),
\label{eq:c7}
\ee
while the spin-orbit interaction couples to the difference of the $\hat{y}$-components of the currents, 
\be
{\cal V}_{\rm so} = 2\alpha_R k_F \int dx~ (J^y_R - J^y_L).
\label{eq:c8}
\ee
Using again extended $SU(2) \times SU(2)$ symmetry  of the non-interacting spin Hamiltonian with respect to independent rotations of the right- and left-moving currents, we rotate
spin currents $\vec{J}_R$ and $\vec{J}_L$ about the $\hat{z}$-axis in opposite directions so as to bring
``vectorial" perturbation  ${\cal V} = {\cal V}_{\rm so} + {\cal V}_{\rm x}$ into the standard Zeeman form, with {\em total} field $h= \sqrt{b^2 + (2\alpha_R k_F)^2}$
along the $\hat{x}$-axis
\be
{\cal V} = - h \int dx~ (M^x_R +  M^x_L).
\label{eq:c9}
\ee
Compare this with \eqref{eq:9} where the field $h$ is pointing along the $\hat{z}$-axis.

The required chiral rotation is given by 
\be
\vec{J}_R = {\cal R}_z(\beta_R) \vec{M}_R,
\quad
\vec{J}_L = {\cal R}_z(\beta_L) \vec{M}_L,
\label{Rotation}
\ee
where the rotation matrix is ${\cal R}$ 
\bea
{\cal R}_z(\beta) =
\left(
\begin{array}{ccc}
\cos\beta & \sin\beta & 0\\
-\sin\beta & \cos\beta & 0\\
0 & 0 & 1\\
\end{array}
\right).
\label{eq:rot}
\eea
The rotation angles are given by 
\be
\beta_R = -\beta_L = \beta= \arctan(2\alpha_R k_F/b).
\label{eq:a10}
\ee
These rotations do not affect ${\cal H}_\sigma^0$ \eqref{eq:6}, which retains its form in the rotated $M$-basis
\be
{\cal H}^0_{\rm \sigma} = \frac{2\pi v_F}{3} \sum_{a=x,y,z} \int dx~  (M^a_R M^a_R + M^a_L M^a_L).
\label{eq:10}
\ee

In terms of the right- and left-moving fermions, the rotation \eqref{eq:rot} corresponds to the rotation of spinors 
$\Psi_{R/L} = (\Psi_{R/L \uparrow}, \Psi_{R/L \downarrow})^T$,
\be
\Psi_{R} = e^{i \beta \sigma^z/2} \Psi'_{R} ,
\quad
\Psi_{L} = e^{-i \beta \sigma^z/2} \Psi'_{L}.
\label{eq:rot-fermions}
\ee

As before, the charge currents \eqref{eq:charge-current1} do not transform under the rotations \eqref{Rotation} and \eqref{eq:rot-fermions} -- the Hamiltonian of the charge sector 
${\cal H}^0_\rho + {\cal H}_{\rm int, \rho}$ is not affected.
The new (primed) fermions parameterize the rotated currents $\vec{M}_{R/L}$
in the same way as the old (unprimed) ones parameterize the currents $\vec{J}_{R/L}$.
For example, under the right rotation ${\cal R}_z(\beta)$
\be
\vec{J}_R = 
\, :\! \Psi^\dagger_{R} \frac{\vec{\sigma}}{2} \Psi_{R}\!: \,\to
\vec{M}_R = 
\, :\!\Psi'^\dagger_{R} \frac{\vec{\sigma}}{2} \Psi'_{R}\!: .
\label{eq:11}
\ee

The interaction in the spin sector ${\cal H}_{\rm int, \sigma}$ \eqref{eq:int-spin}, is strongly modified by the rotation and changes to
\bea
{\cal H}_{\rm int, \sigma} &=& - g \int dx~ \vec{M}_R {\cal R}^T(\beta_R) {\cal R}(-\beta_R) \vec{M}_L \nonumber\\
&=& - g\int dx \left[
M^z_R M^z_L+
 \cos\chi (M^y_R M^y_L + M^x_R M^x_L)
\right.
\nonumber\\
&&\left.\qquad\qquad
+ \sin\chi (M^y_R M^x_L - M^x_R M^y_L) \right],
\label{eq:12}
\eea
where $\chi=\beta_R - \beta_L = 2\beta$ is the {\em relative} rotation angle.

The net field $h$, \eqref{eq:c9}, pointing along the $\hat{x}$-axis, induces incommensurate fluctuations in the system which make some of the terms in 
\eqref{eq:12} to oscillate fast with the coordinate. 
To account for this important effect we proceed as follows: 

1) do a {\em global} rotation of $\vec{M}_{R/L}$ about the $\hat{y}$-axis in order to make external field $h$ \eqref{eq:c9} to point along the $\hat{z}$ axis. 
This is achieved by the following transformation to the new $L$-basis,
$(M^x, M^y, M^z)^T = {\cal R}_y(\pi/2) (L^x, L^y, L^z)^T = (L^z, L^y, -L^x)^T$.
The corresponding rotation for fermions reads $\Psi'_{R/L} \to e^{-i \pi \sigma^y/4} \Psi''_{R/L}$. Here, similar to \eqref{eq:11}, 
$L^a_{R/L} = \frac{1}{2} :\!\Psi''^\dagger_{R/L} \sigma^a \Psi''_{R/L}\! : \,$.

Non-interacting Hamiltonian \eqref{eq:5} is invariant under constant-angle rotations \eqref{eq:rot-fermions} and $\Psi'_{R/L s} \to e^{i \pi \sigma^y/4} \Psi''_{R/L s}$, while the field-dependent term \eqref{eq:c9} is rotated into 
${\cal V} = - h \int dx~ (L^z_R +  L^z_L)$. It is then easy to see that $h$ can be absorbed into fermions $\Psi''_{R/L}$ by a simple $x$-dependent transformation
\be
\Psi''_{R} \to e^{i t_\varphi x \sigma^z/2} \Psi''_{R},
\quad
\Psi''_{L} \to e^{-i t_\varphi x \sigma^z/2} \Psi''_{L},
\quad
t_\varphi = h/v_F,
\label{eq:a14}
\ee
under which kinetic energy \eqref{eq:5} transforms into that of rotated $\Psi''$ fermions {\em plus} $\int dx~ h (L^z_R +  L^z_L)$ term which exactly compensates the
rotated ${\cal V}$ one.

2) 
As a result of this shift the transverse components $L^x_{R/L} \pm i L^y_{R/L} = L^\pm_{R/L}$ of the rotated spin current acquire oscillating position-dependent factors,
$L^+_R \to L^+_R e^{-i t_\varphi x}, L^+_L \to L^+_L e^{i t_\varphi x}$. 
The immediate consequence of this is that many terms in ${\cal H}_{\rm int, \sigma}$ \eqref{eq:12} acquire $x$-dependent oscillations,
\bea
{\cal H}_{\rm int, \sigma} &=& 
- g \!\int\! dx \!\left\{ \cos\chi L^z_R L^z_L 
 - \frac{\cos\chi - 1}{4} 
(L^+_R L^+_L + {\text{h.c.}}) \right.\nonumber\\
&&\qquad\qquad
+ \frac{\cos\chi + 1}{4} (L^+_R L^-_L e^{- i 2t_\varphi x} + {\text{h.c.}}) \nonumber\\
&&\left.\qquad\qquad\!
- i \frac{\sin\chi}{2}\!\left[(L^z_R L^+_L + L^-_R L^z_L)e^{i t_\varphi x}
    - {\text{h.c.}}\right]\! \right\}\! .
\nonumber\\
\label{eq:16}
\eea
Provided that the running backscattering coupling constant $g/v_F$ is small, all
oscillating terms, which represent momentum-nonconserving two-particle
scattering processes, average out to zero. Assuming this, we are allowed to drop all oscillating terms in \eqref{eq:16}. 

The meaning of \eqref{eq:a14} is simple. 
It represents splitting of the Fermi-momentum $k_F$ into the spin-dependent ones $k_{F s} = k_F + s t_\varphi/2$. 
Given that $k_F$ is determined by the particle density, $k_F = \pi N_s^0/L$, the development of the spin-dependent Fermi momenta $k_{F s} = \pi N_s/L$
describes the appearance of the finite magnetization with $N_\uparrow > N_\downarrow$ in the {\em magnetized} ground state of the rotated system. 
Therefore, $\Delta k_F = t_\varphi/2 = \pi (N_\uparrow - N^0_\uparrow)/L$, so that $t_\varphi L = 2 \pi (N_\uparrow - N^0_\uparrow) = - 2 \pi (N_\downarrow - N^0_\downarrow)$ is an integer multiple of $2\pi$
since $N_s$ and $N^0_s$ are integers describing number of spin-$s$ electrons in the system with finite $h\neq 0$ and with zero $h=0$, correspondingly.

3) Having absorbed the $h$-field \eqref{eq:c9} in the preceding step, we now apply global rotation back, by $-\pi/2$ about the $\hat{y}$-axis, to the non-oscillating terms (first line) in \eqref{eq:16}.
So that $(L^x, L^y, L^z)^T = {\cal R}_y(-\pi/2)(K^x, K^y, K^z)^T = (-K^z, K^y, K^x)^T$ and we obtain non-oscillating 
part of the spin-interaction Hamiltonian to be
\bea
{\cal H}_{\rm int, \sigma} 
&=& - \int_0^L dx \left[g_x K^x_R K^x_L  + g_c (K^z_R K^z_L - K^y_R K^y_L) \right] \nonumber\\
&=& - \int_0^L dx \, g_a K^a_R K^a_L ,
\label{eq:ca3}
\eea
where $g_x = g \cos\chi$, $g_z = - g_y = g_c = g (1-\cos\chi)/2$.
Here the fermions rotate as $\Psi''_{R/L} \to e^{i \pi \sigma^y/4} \psi_{R/L}$. Note close similarity of \eqref{eq:ca3} with \eqref{eq:a3} as well as the fact that the
roles of $\hat{x}$ and $\hat{z}$ axes are interchanged in these two expressions.

Under steps 1-3 the non-interacting spin Hamiltonian \eqref{eq:10} transforms into that in terms of spin currents $K^a_{R/L}$,
\be
{\cal H}^0_{\rm \sigma} = \frac{2\pi v_F}{3} \sum_{a=x,y,z} \int_0^L dx~  (K^a_R K^a_R + K^a_L K^a_L).
\label{eq:ca10}
\ee

At this stage the complete Hamiltonian of the spin sector is given by the sum of equations \eqref{eq:ca10} and \eqref{eq:ca3}.
The magnetic field is absent from the above Hamiltonian because it is absorbed into renormalization of the Fermi momenta $k_F \to k_{F s}$.

Tracing the above steps 1-3 we find relation between $\Psi'_{R/L}$ and rotated fermions $\psi_{R/L}$, in terms of which the spin Hamiltonian, \eqref{eq:ca10} and \eqref{eq:ca3},
and the charge Hamiltonian, ${\cal H}^0_{\rm \rho}$ in \eqref{eq:6} and ${\cal H}_{\rm int, \rho}$ in \eqref{eq:int-charge}, are now formulated,
\bea
\label{eq:a12}
&&\Psi'_{R}(x) = e^{-i \pi \sigma^y/4} e^{i t_\varphi x \sigma^z/2} e^{i \pi \sigma^y/4} \psi_{R} = {\cal A}(x) \psi_{R}(x),\nonumber\\
&&\Psi'_{L}(x) = e^{-i \pi \sigma^y/4} e^{-i t_\varphi x \sigma^z/2} e^{i \pi \sigma^y/4} \psi_{L} = {\cal A}(-x) \psi_{L}(x),\nonumber\\
&&{\cal A}(x) = \sigma^0  \cos(\frac{t_\varphi x}{2}) + i \sigma^x \sin(\frac{t_\varphi x}{2}) . 
\eea

We are now in position to understand the boundary condition for the rotated fermions. For the original fermions the open
boundary requires that $\Psi_s(x=0)=0 = \Psi_s(x=L)$, which means that their right- and left-moving components are related as 
\be
\Psi_{R s}(0) = - \Psi_{L s}(0), \quad
\Psi_{R s}(L) = - \Psi_{L s}(L)
\label{o22}
\ee
After the chiral rotation \eqref{eq:rot-fermions} fermions $\Psi'_{R/L}$ obey
\be
\Psi'_{L s}(0) = - e^{i s \beta} \Psi'_{R s}(0), \quad
\Psi'_{L s}(L) = - e^{i s \beta} \Psi'_{R s}(L),
\label{eq:a6}
\ee
where $s = +1$ for the up-spin and $s=-1$ for the down-spin. In matrix notations, \eqref{eq:a6} is just $\Psi'_{L}(x_o) = - e^{i \beta \sigma^z} \Psi'_{R}(x_o)$,
where $x_o=0, L$ denotes wire's open ends.

Next, Eq.\ \eqref{eq:a12} shows how $\Psi'_{R/L}(x)$ transform as a result of global rotations in steps 1-3.
Therefore the boundary condition \eqref{eq:a6} actually reads ${\cal A}(-x_o) \psi_{L}(x_o) = - e^{i \beta \sigma^z} {\cal A}(x_o) \psi_{R}(x_o)$.
Observing that 
${\cal A}^{-1}(x_o) = {\cal A}(-x_o)$, we get 
\bea
\label{eq:a13}
&&
\psi_{L}(x_o) = - {\cal A}(x_o) e^{i \beta \sigma^z} \! {\cal A}(x_o) \psi_{R }(x_o) 
= -\mathcal{B}\psi_R(x_o), \\
&&
\mathcal{B}
= 
\begin{pmatrix}
\cos (t_\varphi x_o) \cos\beta + i \sin\beta& i\cos\beta \sin (t_\varphi x_o)\\
i\cos\beta \sin (t_\varphi x_o) & \cos (t_\varphi x_o) \cos\beta - i \sin\beta\\
\end{pmatrix} \!.
\nonumber
\eea
The matrix $\mathcal{B}$ reduces to $e^{i\beta \sigma^z}$ when $t_\varphi x_o = \pi (N_\uparrow - N_\downarrow)=2\pi M$, as discussed below \eqref{eq:16}. 
Hence at the end of the day \eqref{eq:a13} leads to 
\be
\psi_{L}(x_o) =  - e^{i \beta \sigma^z}  \psi_{R}(x_o),
\label{eq:ca19}
\ee
so that boundary conditions for spinors $\psi_{R/L}$ coincides with those for $\Psi'_{R/L}$, see \eqref{eq:a6}.

Observe that by construction spin current operators $K^a_{R/L}$  in \eqref{eq:ca3} and \eqref{eq:ca10} are given by 
\be
K^a_{R/L} =\, : \!\psi_{R/L}^\dagger \frac{\sigma^a}{2} \psi_{R/L}\! : .
\label{eq:a20}
\ee
Relation between $M^a_{R/L}$ and $K^a_{R/L}$ currents is established with the help of equations \eqref{eq:11}, \eqref{eq:a12} and \eqref{eq:a20},
\be
M^a_R(x) =\, : \!\psi_R^\dagger(x) {\cal A}^\dagger(x) \frac{\sigma^a}{2}
 {\cal A}(x)\psi_R(x) \! : .
\label{eq:ca21}
\ee
We find that they are connected by a $x$-dependent rotation about the $\hat{x}$-axis,
\bea
\vec{M}_R &=& {\cal R}_x(-t_\varphi x) \vec{K}_R =
\left(
\begin{array}{ccc}
1 & 0 & 0\\
0 & \cos (t_\varphi x) & \sin (t_\varphi x) \\
0 & -\sin (t_\varphi x) & \cos (t_\varphi x) \\
\end{array}
\right) \! \vec{K}_R, \nonumber\\
\vec{M}_L &=& {\cal R}_x(t_\varphi x) \vec{K}_L .
\label{eq:ca22}
\eea
Since at the boundary $e^{i t_\varphi x_o} = 1$, we find that there the two operators coincide, $M^a_{R/L}(x_o) = K^a_{R/L}(x_o)$.

In terms of the original spin currents the OBC $\vec{J}_R(x_o)=\vec{J}_L(x_o)$ becomes 
$\vec{M}_R(x_o) = {\cal R}_z(-2\beta) \vec{M}_L(x_o)$, which means that
$M^z_R(x_o) = M^z_L(x_o), M^+_R(x_o) = e^{i 2\beta} M^+_L(x_o)$.
Given the relation $M^a_{R/L}(x_o) = K^a_{R/L}(x_o)$ derived above, we obtain that at the open boundaries the currents $K^a_{R/L}$ obey the same boundary condition as $M^a_{R/L}$
\be
K^z_R(x_o) = K^z_L(x_o),
\quad
K^+_R(x_o) = e^{i 2\beta} K^+_L(x_o),
\label{eq:ca1}
\ee
and, moreover, at $x_o = 0, L$ the original currents $J^a_{R/L}$ and $K^a_{R/L}$ are connected by chiral rotations \eqref{eq:rot}
\be
\begin{split}
&\vec{J}_R(x_o) = {\cal R}_z(\beta_R) \vec{K}_R(x_o),\\
&\vec{J}_L(x_o) = {\cal R}_z(\beta_L) \vec{K}_L(x_o).
\end{split}
\label{c-Rotation}
\ee

\subsection{Hamiltonian and the RG analysis}
\label{app:Ham}

We are now ready to write down the Hamiltonian of the wire of finite length $L$ with open boundaries at $x_o = 0, L$. The simplest way to derive the free part of the Hamiltonian
is to go back to the original fermion formulation, equations \eqref{eq:5}, \eqref{eq:9} and \eqref{eq:rot-fermions}, and observe that rotation \eqref{eq:rot-fermions} leaves \eqref{eq:5} invariant.
The same is not true for the $x$-dependent rotation \eqref{eq:a12} which, in addition to the kinetic energy of $\psi_{R/L}$ fermions, produces the {\em opposite} of \eqref{eq:9} so as to cancel
the field $h$ term  \eqref{eq:9} completely. This, of course, is exactly the purpose of the steps 1-3 and transformation \eqref{eq:a12} as explained in Appendix~\ref{app:Rot}.
In this way we arrive at ${\cal H}^0_{\rm \sigma}$ in \eqref{eq:ca10}.

It is useful to remark here that there is another, slightly more involved way to derive this result is to start with equation \eqref{eq:10} and apply rotations \eqref{eq:ca22} to it. Doing so requires one to implement
a careful point-splitting procedure and treat $(M^a_R)^2$ as a limit of $M^a_R(x) M^a_R(y)$, with subsequent limit $x\to y$ at the end of the calculation. Then, using operator product expansion
(OPE) of SU(2) currents and fermion bilinears \cite{gogolin2004}
\be
\begin{split}
\psi_{R/L s}^\dagger(x) \psi_{R/L s}(y) &= \frac{\pm i}{2\pi (x-y)}
 + :\! \psi_{R/L s}^\dagger(x) \psi_{R/L s}(x) \! : \, , \\
K_{R/L}^a(x) K_{R/L}^b(y) &= -\frac{\delta^{ab}}{8\pi^2 (x-y)^2}
 - \frac{\pm \epsilon^{abc} K^c_{R/L}(x)}{2\pi (x-y)} ,
\end{split}
\label{eq:a24}
\ee
where $x, y$ are spatial coordinates and the limit $x\to y$ is implied. In particular, the first line above helps to establish that the field $h$ produces a constant shift (magnetization) 
of $M^x_R = K^x_R + t_\varphi/4\pi$, which should be added to \eqref{eq:ca22}.
[The same shift of the along-the-field component of the spin current by $h/(4\pi v_F)$ is easily obtained in the abelian bosonization, when one absorbs $-h \partial_x \phi_\sigma$ term by ``completing the square".]
Next, using \eqref{eq:a24} we again arrive at the final result \eqref{eq:ca10} and 
also obtain the cancellation of the $h$-field term \eqref{eq:c9}.

Now we manipulate the interaction term. It is useful to observe that
\bea
g_x K^x_R K^x_L - g_c K^y_R K^y_L
&=& \frac{g_x + g_c}{4} (K^+_R K^+_L + K^-_R K^-_L) \nonumber\\
&&
+\frac{g_x - g_c}{4} (K^+_R K^-_L + K^-_R K^+_L),
\nonumber\\&&
\label{eq:aa4}
\eea
and therefore the interaction Hamiltonian \eqref{eq:ca3} can be written as 
\bea
{\cal H}_{\rm int, \sigma} 
&=& - \!\int_0^L \! dx \!\left[
g_c K^z_R K^z_L   + \frac{g_x + g_c}{4} (K^+_R K^+_L + K^-_R K^-_L)
\right.
\nonumber\\
&& \left.\qquad\qquad
 +\frac{g_x - g_c}{4} (K^+_R K^-_L + K^-_R K^+_L) \right].
\label{eq:ca109}
\eea

Equations \eqref{eq:ca10} and \eqref{eq:ca109} 
represent a non-trivial interacting problem, analysis of which requires renormalization group (RG) treatment. 
The couplings $g_a$ obey the famous BKT RG flow,
\be
\frac{d g_x}{d \ell} = -\frac{g_y g_z}{2\pi v_F} , \quad
\frac{d g_y}{d \ell} = -\frac{g_x g_z}{2\pi v_F} , \quad
\frac{d g_z}{d \ell} = -\frac{g_x g_y}{2\pi v_F} ,
\label{eq:19}
\ee
where $\ell=\log(\alpha'/\alpha)$ describes increase of the short-distance cutoff from $\alpha$ to $\alpha'$.
As discussed in detail in \cite{chan2017}, the solution to the RG equations \eqref{eq:19}
depends on the initial values of the couplings involved, 
\be
g_x(0) = g \cos\chi, \quad
g_z(0) = - g_y(0) = g_c = \frac{g}{2} (1-\cos\chi).
\label{eq:a15}
\ee
Noting that $d(g_y^2 - g_z^2)/d\ell =0$ and the fact that for $\ell=0$ $g_z + g_y = 0$, we conclude that $g_z(\ell) = - g_y(\ell) = g_c(\ell)$ for all $\ell$.
Equations \eqref{eq:19} then reduce to the two coupled equations
\be
\frac{d g_x}{d \ell} = \frac{g_c^2}{2\pi v_F} ,
\qquad
\frac{d g_c}{d \ell} = \frac{g_c g_x}{2\pi v_F} ,
\label{eq:a16}
\ee
which too is characterized by the integral of motion $Y = g_x^2(\ell) - g_c^2(\ell)$.

In the case of comparable spin-orbit and Zeeman energies ($\cos\chi \approx 1/3$), which is the focus of this paper, the combination $g_x + g_c$ towards positive infinity.
This describes development of the correlated SDW state. This means that the combination $g_x - g_c = Y/(g_x + g_c)$ flows to zero in the same limit.
As a result, \eqref{eq:ca109} simplifies to 
\be
{\cal H}_{\rm int, \sigma} =  
- \!\int_0^L \! dx \!\left[
g_c K^z_R K^z_L   + \frac{g_x + g_c}{4} (K^+_R K^+_L + K^-_R K^-_L)
\right] \! .
\label{eq:ca110}
\ee
Equations \eqref{eq:ca10} and \eqref{eq:ca110} constitute the basis for the subsequent analysis. 

\subsection{Bosonization}
\label{app:Bos}

Boundary conditions \eqref{eq:ca19} represent only a slight modification of the OBC considered in Appendix \ref{app:Bos-OBC}. 
They are satisfied by the following representation of the fermion operators \cite{Eggert1992,Wong1994}
\begin{subequations}
\label{eq:a7}
\bea
\psi_{R s}(x) &=&
\frac{i e^{- i s \beta/2}}{\sqrt{2\pi \alpha}} \eta_s 
e^{-i \tilde{\theta}_s} e^{i \frac{\pi x}{L} N_s} e^{i \sqrt{4\pi} \Phi_{R s}(x)}, \\
\psi_{L s}(x) &=& 
\frac{-i e^{i s \beta/2}}{\sqrt{2\pi \alpha}} \eta_s 
e^{-i \tilde{\theta}_s} e^{-i \frac{\pi x}{L} N_s} e^{i \sqrt{4\pi} \Phi_{R s}(-x)}, \qquad\\
\Phi_{R s}(x) &=& 
\sum_{n=1}^\infty \frac{e^{-\alpha q_n/2}}{\sqrt{4\pi n}}
 (e^{i q_n x} b_{n s} + e^{-i q_n x} b_{n s}^\dagger), 
\eea
\end{subequations}
where $q_n = \frac{\pi n}{L}$, $b_{n s}$ is canonical boson with $[b_{n s}, b_{m s'}] = \delta_{n,m} \delta_{s,s'}$, $[\tilde{\theta}_s, N_s]=i$,
$\eta_s$ is the Majorana Klein factor satisfying $\{\eta_s, \eta_{s'}\} = 2 \delta_{s,s'}$, and $N_s$ is the (integer) 
number of particles relative to the equilibrium $N_0$ value. 
Note that following the constructive bosonization \cite{delft1998}, $\eta_s e^{-i \tilde{\theta}_s} = F_s$ is the fermion number-changing operator, $[N_s, F_s] = -F_s$ and
$F_s^\dagger F_s = 1$.
Also notice that $\Phi_{R s}(x)$ is $2L$-periodic.

As usual, we define commuting charge $\Phi_{R \rho}$ and spin $\Phi_{R \sigma}$ bosons 
\be
\Phi_{R \rho} = \frac{1}{\sqrt2}(\Phi_{R \uparrow} + \Phi_{R \downarrow}), \quad
\Phi_{R \sigma} = \frac{1}{\sqrt2}(\Phi_{R \uparrow}-\Phi_{R \downarrow}).
\label{eq:a34}
\ee

Observe that \eqref{eq:a7} implies that in fact 
\be
\psi_{L s}(x) = - e^{i s \beta} \psi_{R s}(-x) 
\label{eq:a8}
\ee
for {\em all} $x \in [0,L]$, and not only for the wire's end-points $x_o=0, L$ in \eqref{eq:ca19}. This is a very general consequence of the chiral nature of one-dimensional fermions, 
see for example Fabrizio-Gogolin formulation \cite{fabrizio1995} of the OBC.

Using bosonization \eqref{eq:a7} we obtain
\bea
K_R^+(x) &=& \frac{e^{i\beta}}{2\pi \alpha} e^{-i \frac{\pi x}{L}} {\cal F}^\dagger e^{-i \frac{2\pi x}{L} M} e^{-i\sqrt{8\pi} \Phi_{R \sigma}(x)},\nonumber\\
K^z_R(x) &=& \frac{M}{2L} + \frac{1}{\sqrt{2\pi}} \partial_x \Phi_{R \sigma}(x),
\quad
K^z_L(x) = K^z_R(-x), \nonumber\\
K_L^+(x) &=& e^{-i 2\beta} K_R^+(-x),
\label{eq:a9}
\eea
where $M = N_\sigma=(N_\uparrow - N_\downarrow)/2$ is the magnetization operator, $\Phi_{R \sigma} = (\Phi_{R \uparrow}-\Phi_{R \downarrow})/\sqrt{2}$ is the spin boson, 
and we used $e^{-i \frac{\pi x}{L}N_s} F^\dagger_s = F^\dagger_s e^{-i \frac{\pi x}{L}(N_s+1)}$.

\subsection{Un-folding of the spin Hamiltonian}
\label{app:unfold}

Next, relation \eqref{eq:a9} allows us to write $[\vec{K}_L(x)]^2$ as $[\vec{K}_R(-x)]^2$, so that $\int_0^L dx [\vec{K}_R(-x)]^2 = \int_{-L}^0 dx [\vec{K}_R(x)]^2$ and 
\eqref{eq:ca10} can be {\em un-folded} onto $(-L,L)$ interval as
\be
{\cal H}^0_{\rm \sigma} = \frac{2\pi v_F}{3} \int_{-L}^L dx \, [\vec{K}_R(x)]^2  
= 2\pi v_F \int_{-L}^L dx \, [K^z_R(x)]^2 .
\label{eq:a24}
\ee
The interaction part \eqref{eq:ca110} can be written, with the help of \eqref{eq:a9}, as 
\bea
{\cal H}_{\rm int, \sigma} &=& 
- \int_{-L}^L \frac{dx}{2} \Big\{ g_c K^z_R(x) K^z_R(-x) \nonumber\\
&&\qquad\qquad
+ \frac{g_x + g_c}{4}[e^{-i 2\beta} K^+_R(x) K^+_R(-x) + {\rm h.c.}]\Big\}.
\nonumber\\&&
\label{eq:a26}
\eea
Equations \eqref{eq:a24} and \eqref{eq:a26} constitute complete spin Hamiltonian of the open quantum wire, written in terms of the chiral (right) current $\vec{K}_R$.
It is worth adding here that charge currents \eqref{eq:charge-current1} and charge Hamiltonian ${\cal H}^0_{\rm \rho} + {\cal H}_{\rm int, \rho}$ are not affected by the rotations.

\subsection{Observables}
\label{app:observe}

Here we express spin operators in terms of terms of boson field $\tilde{\Phi}$ \eqref{eq:a27}. Uniform spin current is easy, using \eqref{eq:a9} and setting $K_\sigma=2$,
\bea
K_R^+(x) &=& \frac{e^{i\beta}}{2\pi \alpha} e^{-i \frac{\pi x}{L}} {\cal F}^\dagger e^{-i \frac{2\pi x}{L} M} \nonumber\\
&&\times e^{-i\sqrt{4\pi}  [ \tilde{\Phi}(x) - \tilde{\Phi}(-x)] - i \sqrt{\pi}  [ \tilde{\Phi}(x) + \tilde{\Phi}(-x)] }.
\quad
\label{x2}
\eea
Observe that it does not contain charge fields. The original spin currents $\vec{J}_{R/L}$ and $\vec{K}_{R/L}$ are connected by \eqref{Rotation} and \eqref{eq:ca22}.

The $2k_F$-component of the spin density \eqref{x1} requires more work. First of all, by \eqref{eq:rot-fermions} and \eqref{eq:a12}
\bea
N^+(x) &=& 
\frac{1}{2}\Psi'^\dagger_R(x)(\sigma^x + i \sigma^y)\Psi'^{}_L(x) \nonumber\\
&=& \frac{1}{2} \psi_R^\dagger(x) {\cal A}(-x) (\sigma^x + i \sigma^y) {\cal A}(-x) \psi_{L}(x).
\nonumber\\&&
\label{x3}
\eea
This gives
\be
N^+(x) = \widetilde{N}^+(x) - i \sin(t_\varphi x) \widetilde{N}^0(x) 
+ [\cos(t_\varphi x)-1] \widetilde{N}^x(x),
\label{x4}
\ee
where $ \widetilde{N}^+ = \psi_{R \uparrow}^\dagger \psi_{L \downarrow}$ reads
\bea
\widetilde{N}^+(x) &=& -\frac{1}{2\pi \alpha} e^{- i\frac{\pi x}{L}(N_\rho +1)} e^{-i \sqrt{2\pi} [\Phi_{R \rho}(x) - \Phi_{R \rho}(-x)]} \nonumber\\
&&\times e^{i \frac{\pi x}{L} M} e^{-i \sqrt{\pi} [\tilde{\Phi}(x) + \tilde{\Phi}(-x)]} e^{-i \frac{\pi x}{L} M} {\cal F}^\dagger .
\label{x5}
\eea
Here we defined
$\widetilde{N}^{a} =  \frac{1}{2}\psi^{\dagger}_R \sigma^a \psi_L$ ($a=0,x,y,z$), and $\sigma^0$ denotes the identity matrix. 
The last three factors in the above equation combine into $f^\dagger(x)$ operator, see \eqref{eq:32}.
We see that $\widetilde{N}^{x,y}$ depends on symmetric combination of the spin modes $\tilde{\Phi}(x) + \tilde{\Phi}(-x)$, similar to \eqref{x5}, while
$\widetilde{N}^{0,z}$ depends on the antisymmetric one $\tilde{\Phi}(x) - \tilde{\Phi}(-x)$. Also important is that \eqref{x5} and other components of $\widetilde{N}^a$ 
depend also on the critical charge mode via the antisymmetric charge combination $\Phi_{R \rho}(x) - \Phi_{R \rho}(-x)$.

Moreover, it is easy to see that the potential part of $\widetilde{H}_\sigma$ [the first line of \eqref{tilde H_sigma}] can be written as 
\be
\widetilde{H}_\sigma \propto \frac{g_x + g_c}{8} \Big(\widetilde{N}^+(x)  \widetilde{N}^+(-x) + {\rm h.c.}\Big) .
\label{xx5}
\ee
Therefore ${\widetilde{\cal H}}_\sigma$ is minimized when $\widetilde{N}^+(x)  \widetilde{N}^+(-x) = -1$, which means that the spin part of $ \widetilde{N}^+(x)$ is reduced 
$\pm i$. That is, 
\be 
\widetilde{N}^+(x) \to \frac{\pm i}{2\pi \alpha} e^{- i\frac{\pi x}{L}(N_\rho +1)} e^{-i \sqrt{2\pi} [\Phi_{R \rho}(x) - \Phi_{R \rho}(-x)]}.
\label{xxx5}
\ee
Comparison with \eqref{eq:a45.3} shows that similar to the ring geometry case, the open wire situation too is characterized by the finite expectation value of the {\em spin part} of 
$\widetilde{N}^y$ and, correspondingly, zero expectation value for the spin part of $\widetilde{N}^x$. 

\subsection{Charge sector Hamiltonian}
\label{app:obc-charge}

With the help of \eqref{x13} the charge Hamiltonian is given by
\bea
H_\rho &=&
\int_0^L dx \left\{\frac{\pi v_F}{2} [J_R^2(x) + J_L^2(x)] \right.\nonumber\\
&&\left.\qquad\quad + \frac{2U_0 - U_{2K_F}}{4} [J_R(x) + J_L(x)]^2\right\}, \quad
\label{x14}
\eea
where $J_R(x) = \frac{1}{2L}N_\rho+\sqrt{\frac{2}{\pi}} \partial_x \Phi_{R \rho}(x)$ and, in the open wire, $J_L(x) = J_R(-x)$. Therefore \eqref{x14} can be written as 
\bea
H_\rho &=& \int_{-L}^L dx \left\{\left(v_F + \frac{2U_0 - U_{2K_F}}{2\pi}\right) [\partial_x \Phi_{R \rho}(x)]^2 \right.\nonumber\\
&&\left.\qquad\qquad
- \frac{2U_0 - U_{2K_F}}{2\pi} \partial_x \Phi_{R \rho}(x) \partial_x \Phi_{R \rho}(-x) \right\} \nonumber\\
&&\left.{} +\frac{\pi}{4L}\!\left(v_F + \frac{2U_0 - U_{2K_F}}{\pi}\right)\! N_\rho^2 \right.,
\eea
and can be diagonalized similarly to the spin Hamiltonian, see \eqref{eq:31}.
We introduce $\Phi_\rho(x)$ via 
\be
\Phi_{R \rho}(x) = \Phi_\rho(x) \cosh\nu  - \Phi_\rho(-x) \sinh\nu 
\label{x15}
\ee
and find 
\be
H_\rho = \frac{\pi v_\rho}{4L K_\rho} N_\rho^2 + \int_{-L}^L dx \, v_\rho [\partial_x \Phi_\rho(x)]^2
\label{x16}
\ee
provided that $e^{2\nu} = K_\rho$, as given by \eqref{K_rho}, and $v_\rho = v_F/K_\rho$, see \eqref{v_rho}.

Therefore 
\be
\Phi_{R \rho}(x) - \Phi_{R \rho}(-x) = \sqrt{K_\rho} [\Phi_{ \rho}(x) - \Phi_{ \rho}(-x) ]
\ee
and we can evaluate $Q_\rho = e^{-i\sqrt{2\pi}[\Phi_{R \rho}(x) - \Phi_{R \rho}(-x)]}$ from \eqref{eq:a53} by normal ordering it,
\be
\begin{split}
&Q_\rho = \prod_{n=1}^\infty e^{g_n(x) a^\dagger_{\rho, n}} e^{-g_n(x) a_{\rho, n}}
 e^{-\frac{1}{2} g_n^2(x)}, \\
&g_n(x) 
=- \sqrt{\frac{2K_\rho}{n}} e^{-\frac{\pi \alpha n}{2L}} 
\sin\!\left(\frac{\pi x n}{L}\right).
\end{split}
\label{x17}
\ee
We used mode expansion [see \eqref{eq:a7}] 
\be
\Phi_{\rho}(x) = \sum_{n=1}^\infty \frac{e^{-\alpha q_n/2}}{\sqrt{4\pi n}} (e^{i q_n x} a_{\rho, n} + e^{-i q_n x} a_{\rho, n}^\dagger)
\ee
with $q_n = \pi n/L$. We are projecting \eqref{eq:a53} onto the state with no bosons, so that $a_{\rho, n} |\cdot\rangle =0$. The presence of the fermion-number changing operator
${\cal F}$ in $\delta H'$ implies that the perturbation connects states with opposite magnetization parity, $\langle 0| \delta H' |1\rangle \neq 0$. 
Projecting $Q_\rho$ \eqref{x17} onto the states $|0\rangle$ and $|1\rangle$, we find that exponentials of $a_{\rho, n}^\dagger$ and $a_{\rho, n}$ operators reduce to $1$, and 
\bea
Q_\rho = \!\prod_{n=1}^\infty e^{-\frac{1}{2} g_n^2(x)}
= \!\left(\ln\frac{(1 - e^{-\frac{\pi\alpha}{L}})^2}
                     {|1 - e^{-\frac{\pi\alpha}{L}} e^{i\frac{2\pi x}{L}}|}
\right)^\frac{K_\rho}{4}\! .
\eea
For $x=0, L$ it reduces to $1$, corresponding to the case (i) of spin-flip scattering near the open end of the wire.
For $\alpha < x < L$ it gives $Q_\rho = \left\{\pi\alpha/[2 L \sin(\pi x/L)]\right\}^{K_\rho/2}$, which is quoted in the main text, case (ii) in Sec.~\ref{sec:pert}.

\bibliography{majorana1.bib}

\end{document}